\documentclass[11pt,a4paper]{article}
\usepackage{jheppub}
\usepackage{amsmath}
\usepackage{graphicx}
\usepackage{indentfirst}
\usepackage{physics}
\usepackage{verbatim}
\usepackage{longtable}
\usepackage{tikz}
\usetikzlibrary{calc,positioning}
\usetikzlibrary{patterns}

\usepackage{titlesec}
\titleformat*{\section}{\normalfont\Large\bfseries\boldmath}
\titleformat*{\subsection}{\normalfont\large\bfseries\boldmath}

\newcommand{\Tau}{\mathcal{T}}
\numberwithin{equation}{section}
\newcommand{\HirD}[3]{D_{#2}^{(#3)}\left(#1,#1\right)}
\newcommand{\uns}{{\bf u}}
\newcommand{\Uns}{{\bf U}}

\newcommand{\Lm}{z}

\newcommand{\pqwebbody}[9]{
\coordinate (A) at ($#1+(-0.5cm,0.5cm)$);
\coordinate (B) at ($#2+(0.5cm,0.5cm)$);
\coordinate (C) at ($#3+(0.5cm,-0.5cm)$);
\coordinate (D) at ($#4+(-0.5cm,-0.5cm)$);

\coordinate (A') at ($#5+(A)+(-1.2cm,1.2cm)$);
\coordinate (B') at ($#6+(B)+(1.2cm,1.2cm)$);
\coordinate (C') at ($#7+(C)+(1.2cm,-1.2cm)$);
\coordinate (D') at ($#8+(D)+(-1.2cm,-1.2cm)$);

\draw (A) -- (B);
\draw (B) -- (C);
\draw (C) -- (D);
\draw (D) -- (A);

\draw (A) -- (A');
\draw (B) -- (B');
\draw (C) -- (C');
\draw (D) -- (D');
\draw (D') --++ ($(-1cm*#9,-1cm*#9)$);
\draw (D') --++ ($(0*#9,-1cm*#9)$);

}
\newcommand{\pqweb}[9]{\begin{tikzpicture}\pqwebbody{#1}{#2}{#3}{#4}{#5}{#6}{#7}{#8}{#9}\end{tikzpicture}}

\title{
On a 5D UV completion of Argyres-Douglas theories
}

\author{Giulio Bonelli, Pavlo Gavrylenko, Ideal Majtara and Alessandro Tanzini}

\affiliation{SISSA, Via Bonomea 265, 34136 Trieste, Italy}
\affiliation{INFN, Sezione di Trieste, Trieste, Italy}
\affiliation{Institute for Geometry and Physics, IGAP, via Beirut 2, 34136 Trieste, Italy}
    
    \emailAdd{bonelli@sissa.it}
    \emailAdd{pasha.145@gmail.com}
    \emailAdd{imajtara@sissa.it}
    \emailAdd{tanzini@sissa.it}

\abstract{We discuss a novel UV completion of a class of Argyres-Douglas (AD)
theories in the $\Omega$-background by its embedding into the renormalisation group flow from five dimensional $\mathcal{N}=1$ superconformal field theories (SCFT) on $S^1$. This is obtained via analysing these theories in the light of ($q$-)Painlev\'e/gauge theory correspondence, which allows to compute the five dimensional BPS partition functions as an expansion in the Wilson loop vev with integer $q$-polynomials coefficients.
These are derived formulating the gauge theory on a blown-up geometry and using a five-dimensional lift of (topological) operator/state correspondence.
We discuss in detail the phase diagram of the four dimensional limits, pinpointing
the special AD loci. Explicit computations are reported for $\tilde E_1$ SCFT and its limit to H$_0=(A_1,A_2)$ AD theory.}

\begin{document}
\maketitle
\newpage
\section{Introduction}

    The study of strongly coupled quantum field theories (QFTs) is an apical topic in contemporary theoretical physics.
    Actually, while the study of non-supersymmetric theories remains the ultimate goal, the features of networks of QFTs related via RG flow and their geometries are easier to be studied in the sheltered environment of supersymmetric gauge theories, where integrability can be used to tame the intricacies of the quantum corrections.
    
    A particularly interesting set of theories discovered in this framework are the Argyres-Douglas (AD) ones \cite{Argyres:1995jj,Argyres:1995xn}, which are special superconformal theories (SCFT) characterized by mutually non-local light degrees of freedom. A Lagrangian description of these is not presently available, making these theories among the most elusive in the supersymmetric panorama. Finding quantities which are computable along the renormalization group flow which leads to these superconformal points is of great help in elucidating their properties. One important finding in this direction is provided by the supersymmetry enhancement mechanism of \cite{Maruyoshi:2016tqk,Maruyoshi:2016aim,Agarwal:2016pjo,Agarwal:2017roi}, which describes the AD points as special vacua of four dimensional Lagrangian $\mathcal{N}=1$ theories where supersymmetry gets enhanced to $\mathcal{N}=2$, allowing to compute their index via supersymmetric localization. 
   
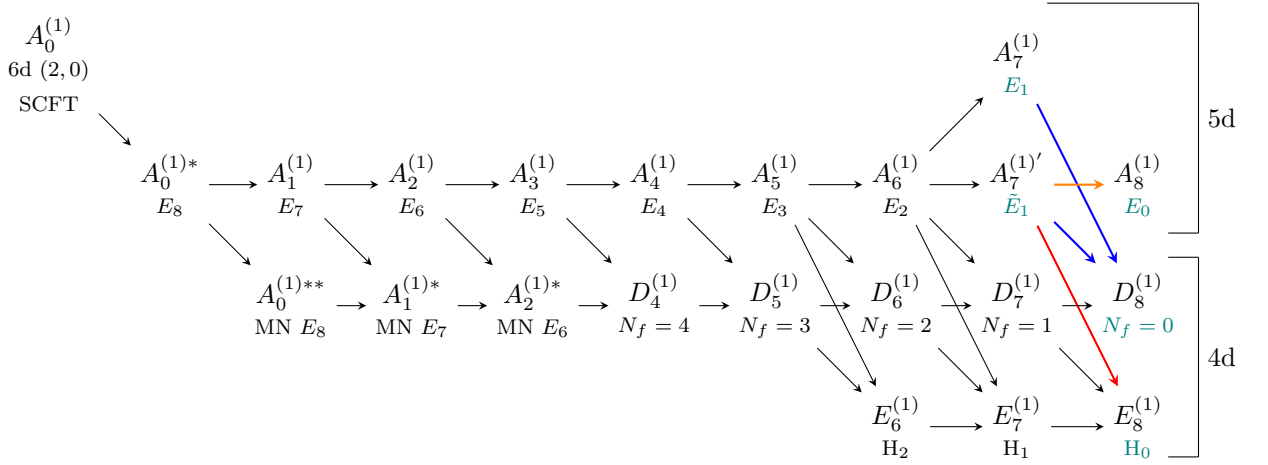
\begin{figure}[t]
\centering\small
 \begin{tikzpicture}[>=stealth,scale=0.8]
 % elliptic/6d
 \node[align=center] (e8e) at (0,4) {$A_{0}^{(1)}$\\[-1.5pt] {\scriptsize 6d $(2,0)$}\\{\scriptsize SCFT}};
 \node[align=center]  (a1qa) at (16,4) {$A_{7}^{(1)}$ \\[-1.5pt] {\scriptsize \color{teal} $E_1$}};
 % q-mult / 5d
 \node[align=center] (e8q) at (2,2) {$A_{0}^{(1)*}$\\[-1.5pt] {\scriptsize $E_8$}};
 \node[align=center] (e7q) at (4,2) {$A_{1}^{(1)}$\\[-1.5pt] {\scriptsize $E_7$}};
 \node[align=center] (e6q) at (6,2) {$A_{2}^{(1)}$\\[-1.5pt] {\scriptsize $E_6$}};
 \node[align=center] (d5q) at (8,2) {$A_{3}^{(1)}$\\[-1.5pt] {\scriptsize $E_5$}};
 \node[align=center] (a4q) at (10,2) {$A_{4}^{(1)}$\\[-1.5pt] {\scriptsize $E_4$}};
 \node[align=center] (a21q) at (12,2) {$A_{5}^{(1)}$\\[-1.5pt] {\scriptsize $E_3$}};
 \node[align=center] (a11q) at (14,2) {$A_{6}^{(1)}$\\[-1.5pt] {\scriptsize $E_2$}};
 \node[align=center] (a1q) at (16,2) {$A_{7}^{(1)'}$\\[-1.5pt] {\scriptsize \color{teal} $\tilde E_1$}};
 \node[align=center]  (a0q) at (18,2) {$A_{8}^{(1)}$\\[-1.5pt] {\scriptsize \color{teal} $E_0$}};
 % diff/q-add/4d lagrangian
 \node[align=center] (e8d) at (4,0) {$A_{0}^{(1)**}$\\[-1.5pt] {\scriptsize MN $E_8$}};
 \node[align=center] (e7d) at (6,0) {$A_{1}^{(1)*}$\\[-1.5pt] {\scriptsize MN $E_7$}};
 \node[align=center] (e6d) at (8,0) {$A_{2}^{(1)*}$\\[-1.5pt] {\scriptsize MN $E_6$}};
 \node[align=center]  (d4d) at (10,0) {$D_{4}^{(1)}$\\[-1.5pt] {\scriptsize $N_f=4$}};
 \node[align=center]  (a3d) at (12,0) {$D_{5}^{(1)}$\\[-1.5pt] {\scriptsize $N_f=3$}};
 \node[align=center]  (a11d) at (14,0) {$D_{6}^{(1)}$\\[-1.5pt] {\scriptsize $N_f=2$}};
 \node[align=center]  (a1d) at (16,0) {$D_{7}^{(1)}$\\[-1.5pt] {\scriptsize $N_f=1$}};
 \node[align=center]  (a0d) at (18,0) {$D_{8}^{(1)}$\\[-1.5pt] {\scriptsize \color{teal} $N_f=0$}};
 % diff/ 4d AD
 \node[align=center]  (a2d) at (14,-2) {$E_{6}^{(1)}$\\[-1.5pt] {\scriptsize H$_2$}};
 \node[align=center]  (a1da) at (16,-2) {$E_{7}^{(1)}$\\[-1.5pt] {\scriptsize H$_1$}};
 \node[align=center] (a0da) at (18,-2) {$E_{8}^{(1)}$\\[-1.5pt] {\scriptsize \color{teal}  H$_0$}};
 % arrows
 \draw[->] (e8e) -> (e8q); \draw[->,blue,thick] (a1qa) -> (a0d);
 \draw[->] (e8q) -> (e7q); \draw[->] (e8q) -> (e8d);
 \draw[->] (e7q) -> (e6q); \draw[->] (e7q) -> (e7d);
 \draw[->] (e6q) -> (d5q); \draw[->] (e6q) -> (e6d);
 \draw[->] (d5q) -> (a4q); \draw[->] (d5q) -> (d4d);
 \draw[->] (a4q) -> (a21q); \draw[->] (a4q) -> (a3d);
 \draw[->] (a21q) -> (a11q); \draw[->] (a21q) -> (a11d); \draw[->] (a21q) -> (a2d);
 \draw[->] (a11q) -> (a1q); \draw[->] (a11q) -> (a1d); \draw[->] (a11q) -> (a1qa); \draw[->] (a11q) -> (a1da);
 \draw[->,orange,thick] (a1q) -> (a0q); \draw[->,blue,thick] (a1q) -> (a0d); \draw[->,red,thick] (a1q) -> (a0da);
 \draw[->] (e8d) -> (e7d);
 \draw[->] (e7d) -> (e6d);
 \draw[->] (e6d) -> (d4d);
 \draw[->] (d4d) -> (a3d);
 \draw[->] (a3d) -> (a11d); \draw[->] (a3d) -> (a2d);
 \draw[->] (a11d) -> (a1d); \draw[->] (a11d) -> (a1da);
 \draw[->] (a1d) -> (a0d); \draw[->] (a1d) -> (a0da);
 \draw[->] (a2d) -> (a1da); \draw[->] (a1da) -> (a0da);
 \draw ($(a0da)+(0.5,-0.5)$)--++(0.5,0)coordinate (H0s);
 \draw ($(a0d)+(0.5,0.8)$)--++(0.5,0)coordinate (SYMs);
 \draw (H0s)--(SYMs) node[pos=0.5, right] {4d};
 \draw ($(a0q)+(0.5,-0.8)$)--++(0.5,0)coordinate (E0s);
 \draw ($(a1qa)+(0.5,1)$)--++(2.5,0)coordinate (SYM0s);
 \draw (E0s)--(SYM0s) node[pos=0.5, right] {5d};
 \end{tikzpicture}\vspace{-2mm}
\caption{Sakai's classification for Painlev\'e equations by surface type and corresponding susy theories, where we highlighted the ones we study.  The red arrow corresponds to the flow from $\tilde E_1$ SCFT ($q$-PI equation) to 4d AD H$_0=(A_1,A_2)$ theory (PI equation). The blue arrows correspond to the 4d geometric engineering limits and the orange one to the flow to $E_0$ SCFT (local $\mathbb{P}^2$).}\label{fig0:classification-surface}
 \end{figure}

   In this paper, we study a novel class of theories \cite{Bonelli:2020dcp,Closset:2021lhd} whose renormalization group flow includes AD points, by leveraging the relation of five-dimensional supersymmetric gauge theories with $q$-Painlev\'e equations. Our proposal differs from the previous ones as it preserves the full supersymmetry along the whole RG flow. This leads to the possibility of computing the partition function of AD theories from Painlev\'e $\Tau$-functions. 
   The relation with isomonodromic deformation problems has indeed proven to be a very effective tool in this context to globally describe the renormalization group of four and five dimensional supersymmetric gauge theories with
   eight supercharges. One of its incarnations is given by the coincidence of Seiberg's classification of these gauge theories \cite{Seiberg:1996bd} with Sakai's classification\footnote{In the table \ref{fig0:classification-surface} we denote by MN $E_n$ the corresponding Minahan-Nemeschansky 4d SCFTs \cite{Minahan:1996fg}. The latter are related to difference Painlevé equations \cite{boalch2007quivers}.} of Painlev\'e equations \cite{Sakai2001RationalSA}, see figure \ref{fig0:classification-surface}. In this context, our findings lead to a link between the Seiberg's SCFT fixed points in 5d and AD theories through the $q$-Painlev\'e renormalisation group flow. 
   
   The work \cite{Gamayun:2013auu} played a pivotal r\^ole for the detailed explanation of this coincidence in terms of    
   Painlev\'e/gauge theory correspondence further analyzed in \cite{Bonelli:2016qwg,Gavrylenko:2016zlf,Bonelli:2019boe,Bonelli:2019yjd,Bonelli:2021rrg,
Bonelli:2022iob,DelMonte:2022nem,Gavrylenko_2019} and for q-difference equations in \cite{Bershtein:2016aef,Bonelli:2017gdk,Bershtein:2017swf,Bonelli:2020dcp,jimbo2017cftapproachqpainlevevi}.
   A novel approach to this correspondence has been developed in \cite{Bonelli:2024wha}, where topological state/operator correspondence was used to find Hurwitz expansions for the Painlev\'e $\Tau$-function around its zeroes. In particular, it was shown that this solves the gauge theory
   blow-up equations \cite{Nakajima:2003pg} in the Nekrasov-Shatashvili (NS) limit \cite{Nekrasov:2009rc}. We observe that blowup equations can be used to resum the NS twisted superpotential around its poles \cite{Bonelli:2025bmt}.

   In this paper we extend the approach above to $\mathcal{N}=1$ supersymmetric gauge theories on $\mathbb{C}^2\times S^1$ and $q$-Painlev\'e equations. A stracching new result arising from this analysis is an explicit realisation of a new five dimensional UV completion of AD theories in the self-dual $\Omega$-background. The AD point turns out to arise from the perturbation of a finite coupling point of 5d after a proper limit to 4d is taken. We exemplify this for the simplest AD point H$_0$, by showing how this arises from the 5d $SU(2)$ $\mathcal{N}=1$ Super Yang-Mills with Chern-Simons (CS) level $k=1$ at negative coupling. 
   This particular reduction, expected from Sakai's classification, was studied from the view point of the corresponding cluster integrable system in \cite{Bonelli:2020dcp}, from that of its SW geometry in \cite{Closset:2021lhd} and from its BPS spectrum in \cite{DelMonte:2021ytz,DelMonte:2022kxh}. This is a different type of reduction with respect to the combination of sequential mass deformation and circle reduction to 4d studied for example in \cite{Martone:2021drm}. Here, instead, we do a double-scaling limit of the mass parameter and of the circle radius keeping their combination fixed. 
   
   One of the aims of this paper is to make this construction explicit by following the form of the solution of the RG flow. We perform a detailed analysis of the 4d scalings and find among them a set of flows leading to H$_0$ theory, see figure \ref{fig4:phasediagram} for the associated phase diagram.
   This analysis shows how and when the concrete Hurwitz expansion of the AD BPS partition function arises from a suitable continuous limit of the $q$-Painlev\'e $\Tau$-function.
   The latter has an integral \textit{$q$-polynomial Hurwitz expansion}\footnote{By this we mean that the values of the $\Tau$-function in $q$-shifted points, see for example \eqref{sec2,5}, are polynomials with coefficients in the ring $\mathbb{Z}[q]$.} in terms of the vev of the Wilson loop on $S^1$.

   To compute the Hurwitz expansions we first extend the present formulation of the Nakajima-Yoshioka (NY) 5d blowup equations by including the Wilson loop contributions \cite{Wang:2023zcb}. We then take the NS limit of the latter and make use of a 5d version of topological state/operator correspondence to express the $\hat{\mathbb{C}}^2\times S^1_\beta$ partition function (in presence of co-dimension two observables, see formula \eqref{sec1,1}) in terms of polynomials of the NS Wilson loop vev $\Uns$ with $q$-polynomial coefficients, see for example \eqref{sec2,6}, \eqref{sec2,7}, \eqref{sec2,8}. These generalize the Hurwitz expansions for Painlevé equations found in \cite{hone2013properties,hone2017hirota,Bonelli:2024wha} 
   to the $q$-Painlevé case.
   
    For completeness of the analysis we also study the Hurwitz expansion of the $q$-Painlev\'e $\Tau$-function of 5d $SU(2)$ super Yang-Mills with CS level $k=0$ for which we do not find any limit to AD theories.
   This is expected both from Sakai's table and from topological strings considerations.

   As it is well known, $\mathcal{N}=1$ SCFTs on $\mathbb{C}^2\times S^1$ can be geometrically engineered from topological strings on local Calabi-Yau (CY) geometries \cite{Hori:2003ic}. The theories we study in this paper are described by local $\mathbb{F}_{1,0}$ CYs for $k=1,0$ respectively. The limits we study on the $q$-Painlev\'e $\Tau$-functions have a non-trivial geometrical interpretation which confirms our findings about the relation to 4d AD points. In particular, as explained in detail in section \ref{5dsection4}, the AD point H$_0$
is reached by a two steps procedure. One first performs a limit in the local $\mathbb{F}_1$ geometry, leaving behind a one-parameter deformation of local $\mathbb{P}^2$, see figure \ref{fig3:pqweb1}. The AD point is then reached via a further scaling limit to four dimensions which keeps this deformation parameter finite, see  \eqref{sec3,13} and \eqref{sec3,8}. It is now clear why the same point cannot be reached from the local $\mathbb{F}_0$ geometry: in this case there is simply no way to deform the geometry to reach\footnote{This detailed procedure can be better defined in the GLSM description of this geometry and by discussing the analytic continuation
in the FI-parameters before the limits of the negatively coupled theory are taken.} the local $\mathbb{P}^2$.
From the 5d SCFT viewpoint, the one-parameter deformation of local $\mathbb{P}^2$ corresponds to a negative five dimensional gauge coupling and the corresponding geometry describes a negative massive deformation of Seiberg's $\tilde E_1$ SCFT \cite{Aharony:1997bh,Bergman:2013ala,Bergman:2013aca}. It is indeed by performing a suitable 4d limit of the latter that we reach the H$_0$ AD point.

The content of the paper is the following. In section \ref{5dsection2} we review the equivariant topological twist of the 5d gauge theory and the blowup topology changing operator. We discuss the expansion in terms of Wilson loops of the 5d NS blowup partition function and its interpretation in terms of topological operator/state correspondence. In section \ref{5dsection3} we compute explicitly the 5d Hurwitz expansion of the $q$-Painlevé $\Tau$-function for $5d$ SYM theory with CS level $k=0,1$ and we study its SW limit. In section \ref{5dsection4} we discuss two possible 4d limits of the 5d theory and its Hurwitz expansion. The first one is the standard 4d geometric engineering limit and allows to recover the Hurwitz expansion of PIII$_3$ corresponding to the 4d pure gauge theory of \cite{Bonelli:2024wha}. The second limit, which arise in the $k$=1 case only, is the one leading to the Argyres-Douglas SCFT H$_{0}$. Finally, in App.\ref{appA} we display the first coefficients of the Hurwitz expansion studied in section \ref{5dsection3}.

\paragraph{Open problems}
\begin{itemize}
    \item In this paper we lifted the evidence of the Hurwitz ${\mathbb{Z}}$-integrality of the expansion coefficients of Painlev\'e $\Tau$-function around its zeroes\footnote{Strictly speaking this is valid for the odd sector $j=1$. The initial value for the expansion in the even sector $j=0$ is $\Tau=1$ but the integrality properties still hold.} \cite{hone2013properties,hone2017hirota,Bonelli:2024wha} to that of the
    ${\mathbb{Z}}[q]$-integrality of the $q$-Painlev\'e $\Tau$-function. 
    It is still an open problem for us to give a fundamental proof of this fact. 
    Geometrically, this should correspond to ($K$-theoretic) equivariant Donaldson invariants on the blow-up.
    This could be possibly extended to the $\hat{\hat{E_8}}$ theory in 6d on an elliptic curve to be interpreted as an elliptic version of Donaldson theory. We also observe that our expansion takes very simple form at $\Uns=0$, see formula \eqref{new?}. It will be nice to compare our results with the Donaldson-Thomas invariants discussed in \cite{Bridgeland:2024saj}.
    
    \item From Sakai's classification of (q-)Painlev\'e equations one expects that analogue 
    limits can be identified from the 5d $SU(2)$ theories with $N_f$ fundamentals to the AD points of the 4d SU(2) theories with $N_f+1$ fundamentals (see figure \ref{fig0:classification-surface}). In particular, we expect a limit from 5d $N_f=1$ to $H_1$ and $N_f=2$ to $H_2$. Geometrically this should be matched by an appropriate deformation of the corresponding del Pezzo geometries, see \cite{Bonelli:2020dcp,DelMonte:2023vwv} for the $N_f=2$ case and \cite{Closset:2021lhd} for a general description in terms on the classification of rational elliptic curves. 
    
    \item In \cite{Bonelli:2024wha} we used the modular properties of the Hurwitz expansion of the Painlevé $\Tau$-function to derive the holomorphic anomaly equations for the topological string that engineers the gauge theory and to construct non-perturbative background independent completion of the corresponding partition function. We expect that this construction gets naturally lifted to the 5d setting where a deeper understanding of the full generating function of the Wilson loop $\Uns$ is required. This should give the holomorphic anomaly equations of \cite{Wang:2023zcb} and their non-perturbative solutions. An evidence in this direction is the manifest modular invariance of the blowup factor in the SW limit, see formula \eqref{sec3,10.5bis}.  As topological string/spectral theory duality suggests \cite{Grassi:2014uua,Grassi:2014zfa}, the $q$-Painlevé $\Tau$-function can be written as the spectral determinant of the quantum mirror curve. Regarding this point it may be very interesting to give a spectral theory interpretation of the Hurwitz expansion in terms of truncated spectral determinants.   
    
    \item As it turns out from \cite{Moore:1997pc,Manschot:2021qqe} the $SU(2)$ blowup partition function in terms of the IR Coulomb modulus can be derived from the $u$-plane integral of the low-energy theory. This can be promoted to the 5d theory in the form of a $U$-plane integral \cite{Closset:2021lhd} with $U$ being the Wilson loop vev in the SW limit. Our results suggest the existence of an equivariant extension of the above constructions.
 
    \item As we will see in \eqref{sec2,6}, the Hurwitz expansion is given in terms of the IR (quantum) variable $\Uns$. As the 5d Nekrasov partition function has positive radius of convergence in the instanton counting parameter $z$ \cite{Felder:2017rgg}, it would be interesting to see if an alternative expansion of the partition function around the finite coupling point $z_0$ holds. This will be an interesting perspective to further understand the UV completion we propose for the AD theories.
     
    \item The Wilson loop expansion \eqref{sec2,6} turns out to truncate at a maximal degree $n_{\max}(d)$ in $\Uns$, see formulas \eqref{sec3,4.75}, \eqref{sec3,7.75}. 
    In subsection \ref{5dsection2.4} we interpret this
    as a remnant effective contact term in the IR 5d
        theory and in the SW limit we explicitly check this in subsection \ref{SW}. The derivation of this bound from the UV picture is unclear to us. We expect it to be related to the counting of zero modes of the theory living on the codimension 2 defect. 
        
    \item One could ask whether there is a representation theoretic meaning of the Hurwitz Wilson loop expansion and of its truncation. Indeed, it resembles the structure of fusion rules of generalized symmetries, deformed by the presence of a defect
    \begin{equation*}
    1\times1=1 \rightarrow1\times_d1=\sum_{n=0} ^{n_{\max}(d)}B_{d,n}W_n    
    \end{equation*}
    that can be also generalized to the one in presence of Wilson loops \cite{Kim:2021gyj}
    \begin{equation*}
     W_k\times W_l =W_{k+l} \rightarrow W_n\times_d W_m=\sum_{l=0} ^{n_{\max,kl}(d)}(B_d)^n_{kl}W_n \ .   
    \end{equation*}
    In particular, we observe that the structure constants $(B_d)^n_{kl}$ satisfy an associativity property which arises from a double blowup consistency condition.

    \item As we observe in subsection \ref{SW}, special loci of the RG flow arise in the SW limit, including the Seiberg's $E_1$ SCFT. It would be interesting to investigate if they can generate new $K$-theoretic differential invariants for the classification of four-manifolds. Moreover, similar simplification of the flow seem to arise also in presence of $\Omega$-background for $\log q\in2\pi i \mathbb{Q}$. A more detailed analysis is required to understand these points.

    \item In \cite{Manschot:2019pog} the values of the lagrangian $SU(2)$ 4d SCFT central charges $a,c$ are derived from the $\Omega$-background corrections of the prepotential. It is natural to ask if we can compute the central charge $a-c$ of the 4d AD SCFT from the gravitational corrections of the scaled 5d SCFT using \eqref{sec3,8}. These should match those computed in \cite{Xie:2012hs} from compactifying the $(2,0)$ 6D theory on irregular singularities \cite{Bonelli:2011aa}.
    
    \item One could consider to generalize the set-up we discussed in this paper 
    in many ways. For example to higher-rank or quiver gauge theories
    and/or their formulation in an arbitrary $\Omega$-background. The latter 
    is described by a quantum version of $q$-Painlevé equations \cite{Bershtein:2017swf,Bershtein:2018srt}.

\end{itemize}

{\bf Acknowledgments:} 
We would like to thank 
S.~Benvenuti,
M.~Bershtein,
C.~Closset,
F.~Del Monte,
M.~Del Zotto,
A.~Grassi,
J.~Gu,
O.~Lisovyy,
A.~Marshakov,
N.~Nekrasov,
T.~Pedroni,
D.~Rodriguez-Gomez,
A.~Shchechkin,
for useful discussions and comments.

The research of G.B. and I.M. is partly supported by the INFN Iniziativa Specifica ST\&FI and by the PRIN project “Non-perturbative Aspects Of Gauge Theories And Strings”. The research of  P.G. and A.T. is partly supported by the INFN Iniziativa Specifica GAST and Indam GNFM. 
The research is partly supported by the MIUR PRIN Grant 2020KR4KN2 ``String Theory as a bridge between Gauge Theories and Quantum Gravity''.  
All the authors acknowledge funding from the EU project Caligola (HORIZON-MSCA-2021-SE-01; in particular, PG also thanks SCGP, where a part of this work was done), Project ID: 101086123, and CA21109 - COST Action CaLISTA. 

%{\bf Statements and Declarations:} All authors state that there is no conflict of interest, and data sharing is not applicable to this article as no datasets were generated or analysed during the current study.

\section{Five-dimensional gauge theory on the blowup}\label{5dsection2}

\subsection{Topological observables in 5d gauge theory}
The BPS sector of the gauge theory is efficiently described in terms of the topological twist. In this section we will review the equivariant topological twist of 5d $\mathcal{N}=1$ gauge theory and the construction of the topological observables, focusing for simplicity on the pure theory case. 

Consider first 5d SYM theory on flat spacetime $X=\mathbb{C}^2\times S^1_\beta$ compactified on a circle of radius $\beta$. The field content of the theory is given by a 5d vector multiplet $(\varphi, \lambda_\alpha,A_\mu)$ where $\varphi$ is a real scalar, $A_\mu$ is a 5d gauge field and $\lambda_\alpha$ are gauginos. We want to study the theory in the $\Omega$-background $(\epsilon_1,\epsilon_2)$. The theory has an Euclidean spacetime rotation symmetry $SO(4)$ and a $\mathcal{R}$-symmetry coming from the symplectic reality condition for the supercharges. We can then do a partial topological twist replacing $SO(4)\simeq SU(2)_L\times SU(2)_R$ with the twisted rotations $SO(4)'=SU(2)_L\times SU(2)'_R$ where $SU(2)_R'=\text{diag}(SU(2)_R\times SU(2)_\mathcal{R})$. After the topological twist, the field content is $(\varphi,A_\mu,\psi_\mu,b^+_{\mu\nu})$  with $\mu=1,\dots,5$. The field $\psi_\mu$ is a one-form fermion field which is associated to the exterior derivative of $A_\mu$. We need also to introduce an extra bosonic self-dual 2-form field $H^+_{\mu\nu}$ to close the algebra off-shell. In the topological twisted frame the supercharges become $(Q,G_\mu,Q_{\mu\nu}^+)$ and contain a scalar topological charge $Q$. This can be made equivariant combining it with the action of $G_\mu$ and the spacetime symmetry $SO(4)'\times U(1)$ \cite{Nekrasov:2002qd}
\begin{equation}
Q_v=Q+v^\mu G_\mu\ ,   
\end{equation}
where 
\begin{equation}
v=\partial_t+i\sum_{j=1}^2(\epsilon_j z_j\partial_{z_j}-\,\,h.c.)\ ,
\end{equation}
is the vector field associated to the Cartan subalgebra $U(1)\times U(1)\subset SO(4)'$ of spacetime rotations and to the circle translations\footnote{The addition of the circle translations is needed to fully localize the path integral to the susy QM on the instanton moduli space.}. The action of the charge $Q_v$ is then \cite{Nekrasov:1996cz,Qiu:2016dyj}
\begin{align}\label{sec1,0}
&Q_v A = \psi\ , \quad Q_v \psi = \iota_v F + i D \varphi \ ,\quad Q_v \varphi = -i \iota_v\psi\ ,\nonumber\\
&Q_v b^+ = H^+\ , \quad Q_v H^+ = \mathcal{L}^A_v b^++ i [\varphi, b^+]\ , 
\end{align}
where $D$ is the covariant derivative, $\mathcal{L}^A_{v}=D\iota_v+\iota_vD$ is the covariant Lie derivative, and it can be verified that
\begin{equation}
 Q_v^2 = \mathcal{L}^A_{v} +G_{\phi}\ ,
\end{equation}
where $G_{\phi}$ are infinitesimal gauge transformations with respect to the complex scalar $\phi=\varphi+i\iota_v A$. Therefore, the corresponding topological observables are equivariant cohomology classes for the torus action generated by $v$ and which are gauge invariant.

We emphasize that the above twist is a partial topological twist, in the sense that the theory retains a dependence on the circle radius $\beta$ through the instanton action $\beta \int_{M_4}\Tr F\wedge F$. Therefore, the correlation functions of the topological observables given by $Q_v$-cohomology of the 5d theory reduce to the ones of some susy QM on the circle. 
Furthermore, although formally a TQFT, the topological twist associated to $Q_v$ captures richer dynamical informations thanks to equivariance.

The twisted supersymmetry algebra $(Q,G_\mu,Q_{\mu\nu}^+)$ can be generalized to an arbitrary toric 4-manifold $M_4$ which admits a toric action of $U(1)\times U(1)$ and we can extend the above construction to any $X=M_4\times S^1_\beta$.

There are two natural classes of equivariant topological observables which can be defined \cite{Baulieu:1997nj}. The first class is the 5d lift of the 4d local observable $\Tr \phi^2$ and corresponds to the circular Wilson loops placed in the fixed points of the toric action of $M_4$ and wrapped on the compactified circle dimension
\begin{equation}\label{sec1,0.5}
W_R=\Tr_R P\exp\int_{S^1} (\varphi\dd t+iA)=\Tr_R P\exp\int_{S^1} \phi\, \dd t\ ,
\end{equation}
where $\varphi$ is the real scalar in the 5d vector multiplet, $P$ denotes path ordering and $R$ is some representation of the gauge group $G$,
and the associated topological multiplet constructed with the usual descent procedure. 

The second class is constructed observing that the transformations \eqref{sec1,0} can be rewritten as the generalized Bianchi identity for the curvature of the universal bundle \cite{Bershtein:2015xfa}
\begin{equation}
 {\bf DF}=(-Q+D+\iota_v)(F+\psi+i\varphi)=0\ ,
\end{equation}
and for any ad-invariant polynomial $\mathcal{P}({\bf F})$ of the Lie algebra of the gauge group we have
\begin{equation}\label{sec1,0.25}
Q\mathcal{P}({\bf F})=(d+\iota_v)\mathcal{P}({\bf F})\ .
\end{equation}
Using this fact we can easily construct $Q_v$-closed observables as follows. Consider an equivariant cohomology class $\Omega\in H^{\bullet}_v(X)$ then from \eqref{sec1,0.25} the observable
\begin{equation}
O(\Omega,\mathcal{P})=\int_X\Omega\wedge\mathcal{P}({\bf F})\ ,   
\end{equation}
is $Q_v$-closed. In four dimensions for $\mathbb{C}^2$ the natural equivariant cohomology class is the equivariant symplectic form $\omega+H$ where $H$ is the moment map and $\omega$ is the standard symplectic form of $\mathbb{C}^2$
\begin{equation}
H=\frac{\epsilon_1}{2}|z_1|^2+\frac{\epsilon_2}{2}|z_2|^2\ , \quad \omega=\frac{i}{2}\dd z_1\wedge \dd\bar{z}_1+\frac{i}{2}\dd z_2\wedge \dd\bar{z}_2\ , \quad (d+\iota_v)(\omega+H)=0\ .
\end{equation}
This class defines the surface observable
\begin{equation}\label{sec1,0.75}
 O^{(4d)}(\Omega,{\bf F}^2)=\int_{M_4} (\omega+H)\wedge \text{Tr}\,{\bf F}^2=\int_{M_4}\omega\wedge 2\Tr\left(i\varphi F+\frac{1}{2}\psi\wedge\psi\right)+H\Tr F\wedge F\ ,
\end{equation}
For a general toric manifold $M_4$ the same construction applies patch by patch which are then glued together with the appropriate $\epsilon_j$ parameters and sums over fluxes \cite{Bershtein:2015xfa}.

In 5d we still have $(d+\iota_v)(\omega+H)=0$, but the surface observable is naturally lifted to a codimension 2 operator. The lift to 5d of the equivariant symplectic form
in the odd equivariant cohomology is 
\begin{equation}\label{sec1,0.865}
\Omega=(\omega+H)\wedge \dd t-\lambda \ ,   
\end{equation}
where $\lambda=\frac{i}{2}z_1\dd \bar{z}_1+\frac{i}{2}z_2\dd \bar{z}_2$ is the 1-form symplectic potential such that $\omega=d\lambda$ and $\iota_v\lambda=H$. The appearance of the locally defined symplectic potential has the following consequences. 
The 5d lift of the surface observable is
\begin{align}\label{sec1,1}
 &O(\Omega,{\bf F}^2)=\frac{1}{4\pi}\int_X (\omega\wedge \dd t +H\dd t-\lambda)\wedge \text{Tr}\,{\bf F}^2=\\
 &=\frac{1}{4\pi}\int_{X}\ \omega\wedge2\Tr\left(i\varphi F + \frac{1}{2} \psi \wedge \psi \right) \wedge dt + (H\dd t -\lambda) \wedge \text{Tr} \left( F \wedge F \right)=
 \nonumber\\
 &=\frac{1}{4\pi}\int_{X}\ \omega\wedge2\Tr\left(i\varphi F + \frac{1}{2} \psi \wedge \psi \right) \wedge dt-\omega\wedge CS_3(A) + H \, \text{Tr} \left( F \wedge F \right) \wedge dt\ , \nonumber
\end{align}
where we integrated by parts the term in $\lambda$ using $\Tr F\wedge F=\dd CS_3(A)$ with $CS_3(A)$ the CS 3-form
\begin{equation}
 CS_3(A)=\Tr\left[\dd A \wedge A + \frac{2}{3} A\wedge A\wedge A\right]\ ,
\end{equation}
and we normalized by $1/4\pi$ to get the correct normalization for the CS level.
Therefore, in 5d equivariance with respect to circle translations requires the addition of a CS term. This implies that the physical observable must be exponentiated because the CS term is gauge invariant only if it represents an integral cohomology class \cite{Baulieu:1997nj}\footnote{In this paper the form of the susy version of the 5D CS term is also given as
\begin{align*}
&\frac{1}{24 \pi^2}\int_{X} CS_5(A)+Tr[3i\varphi F\wedge F +2\psi\wedge\psi\wedge F]\wedge \dd t  \ , \\
&CS_5(A)=\Tr\left[\dd A \wedge\dd A \wedge A + \frac{3}{2} \dd A \wedge A\wedge A\wedge A+\frac{3}{5} A\wedge A\wedge A\wedge A\wedge A\right] \ . 
\end{align*}
}.
This has the important consequence that the coupling of the exponentiated operator of codimension $2$ is discrete, as it is the analogue of the 3D CS level.

\subsection{Blowup topology changing operator}
Consider the $5d$ Nekrasov partition function on the 5d spacetime $X=\mathbb{C}^2\times S^1_\beta$ of a $SU(2)$ gauge theory with CS level $k$
\begin{equation}
Z^{(k)}(a,\Lm,\epsilon_1,\epsilon_2,\beta)=\ev{1}^{(k)}_{\mathbb{C}^2\times S^1_\beta}\ ,    
\end{equation}
where $\beta$ is the circle radius and $z=\beta^4\Lambda^4$ is the 5d instanton counting scale. From the point of view of radial quantization, cutting a small ``cylinder'' $C_\delta(0)=B^4_\delta(0)\times S^1_\beta$, with the bases identified, and centered in the origin of spacetime, we can intepret the partition function $Z^{(k)}$ as the transition amplitude between an ``in'' state $\ket{C_\delta(0)}$ state given by a path integral on the cylinder and an ``out'' state $\ket{X\setminus C_\delta(0)}$ corresponding to a path integral on the remaining part of spacetime \cite{Nekrasov:2020qcq}
\begin{equation}
Z^{(k)}(a,\Lm,\epsilon_1,\epsilon_2,\beta)= \braket{out}{in}=\braket{X\setminus C_\delta(0)}{C_\delta(0)}\ .  
\end{equation}
Topologically this follows from the identity $\mathbb{C}^2=\mathbb{C}^2\# B^4_\delta(0)$ where $\#$ is the connected sum.

Let us now consider the theory on the manifold $X'=\hat{\mathbb{C}}^2\times S^1_\beta$ where $\hat{\mathbb{C}}^2={\rm tot[}{\mathcal O}(-1)_{{\mathbb P}^1}]$ is the blowup of $\mathbb{C}^2$ at the origin. This introduces a non-trivial equivariant $2$-cycle which is the exceptional divisor $E\simeq\mathbb{C}P^1$. We consider then the codimension 2 defect observable wrapped on $E\times S^1$
\begin{equation}
I(E)\equiv \exp(O(\Omega,{\bf F}^2))\ ,    
\end{equation}
where $\Omega$ is the equivariant version of the Poincaré dual of $E$ in the sense of \eqref{sec1,0.865}. We define now the generating function of $I(E)$ on $X'$ with first Chern class $j$ and CS level $k$
\begin{equation}\label{sec2,0}
\hat Z^{(j,k)}(a,\Lm,\epsilon_1,\epsilon_2,d,\beta) = \ev{ I(E)^{(d+1)}}^{(j,k)}_{\hat{\mathbb{C}}^2\times S^1_\beta}\ .
\end{equation}
The parameter $d\in \mathbb{Z}$ is quantized because from \eqref{sec1,1} it corresponds to the level of the CS term contribution appearing in $I(E)$, and we shifted by $r/2=1$ to make $d$ symmetric with respect to Serre duality $d\to r/2-d$, where $r=2$ is the rank of $U(2)$. 
The partition function $\hat Z^{(j,k)}(d)$ can be expressed in terms of the partition function $Z$ through the Nakajima-Yoshioka (NY) blowup relations (for rank $r=2$) \cite{Nakajima:2005fg,Gottsche:2006bm,Shchechkin:2020ryb}
\begin{align}\label{sec2,1}
\hat Z^{(j,k)}(a,z,\epsilon_1,\epsilon_2,d,\beta) = \nonumber\\
=\sum_{n\in \mathbb{Z}+\frac{j}{2}} e^{(d/4-1/8) \beta(\epsilon_1 + \epsilon_2)} &Z^{(k)}(a+n\epsilon_1,z q_1^{d+k(j-1)/2},\epsilon_1,\epsilon_2-\epsilon_1,\beta)\times\\
&Z^{(k)}(a+n\epsilon_2,z q_2^{d+k(j-1)/2},\epsilon_1-\epsilon_2,\epsilon_2,\beta)\ , \nonumber
\end{align}
where $q_j=e^{\beta \epsilon_j}$. The effect of the observable $I(E)$ is to shift the coupling as $z q_{1,2}^{d+k(j-1)/2}$ in the two patches of $\hat{\mathbb{C}}^2\times S^1_\beta$.

Topologically, the manifold $X'=\hat{\mathbb{C}}^2\times S^1_\beta$ is obtained from the Cartesian product of the circle $S_\beta^1$ times the connected sum $\hat{\mathbb{C}}^2=\mathbb{C}^2\#\overline{\mathbb{C}P^2}$. Therefore, cutting a small cylinder $C_\delta(0)$ at the origin, we can interpret the generating function $\hat Z^{(j,k)}(d)$ as the transition amplitude
\begin{equation}
 \hat Z^{(j,k)}(a,\Lm,\epsilon_1,\epsilon_2,d,\beta)=\braket{X\setminus C_\delta(0)}{\Psi_d} \ , 
\end{equation}
where the ``in'' state $\ket{\Psi_d}$ corresponds to the path integral on $\overline{\mathbb{C}P^2}\times S^1_\beta\setminus C_\delta(0)$ and with $d+1$ insertions of the codimension 2 defect observable $I(E)^{d+1}$ 
\begin{equation}
 \ket{\Psi_d}=I(E)^{(d+1)}\ket{\overline{\mathbb{C}P^2}\times S^1_\beta\setminus  C_\delta(0) }  \ .
\end{equation}

\begin{figure}[htbp]
\centerline{\includegraphics[scale=1]{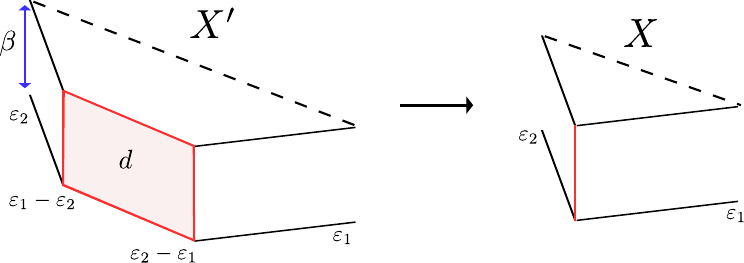}}
\caption[]{\label{fig:5d blowup}Diagram for the blow up $X'=\hat{\mathbb{C}}^2\times S^1_\beta $. The red surface is the support $E\times S^1_\beta$ of the codimension 2 operator $I(E)^{d+1}$. The arrow illustrates the 5d topological operator/state correspondence (blow down).}
\end{figure}

Because $Q_v$-observables are BPS protected we have a 5d version of ``operator/state correspondence'' where each state $\ket{\psi}$ of the Hilbert space defined on the boundary of the cylinder $C_\delta(0)$ can be replaced by some \emph{line} operator supported on the circle in the origin $\{0\}\times S^1_\beta$ , see figure \ref{fig:5d blowup}. 

This implies that all topological observables of the 5d theory can be expanded in the basis of Wilson loops $W_R$ defined in \eqref{sec1,0.5} and by operator/state correspondence the blowup state $\ket{\Psi_d}$ can be expressed as a sum of these Wilson loop operators. We will study the consequences of this in the next sections.

\subsection{NS blowup factor}
We will focus now on the NS limit $(\epsilon_1,\epsilon_2)\to(\epsilon,0)$ of the blowup partition function and we will show that it corresponds to the $q$-Painlevé $\Tau$-function. In order to do this we define the \emph{blowup factor} as the normalized expectation value
\begin{equation}\label{sec2,1.5}
\mathcal{B}^{(j,k)}(a,\Lm,\epsilon_1,\epsilon_2,d,\beta)=\frac{\hat Z^{(j,k)}(a,\Lm,\epsilon_1,\epsilon_2,d,\beta)}{Z^{(k)}(a,\Lm,\epsilon_1,\epsilon_2,\beta)}\ ,
\end{equation}
and we consider the refined genus expansion of the partition function in the $\Omega$-background
\begin{equation}\label{sec2,1.75}
\log Z(a,\Lm,\epsilon_1,\epsilon_2,\beta) \sim \sum_{g=0}^{+\infty}\sum_{k\in \{\frac{1}{2}\}\cup\mathbb{Z}_{\geq0}} (-\epsilon_1\epsilon_2)^{g-1} (\epsilon_1+\epsilon_2)^{2k} \mathcal{F}_{g,k}(a,\Lm,\beta) \ ,
\end{equation}
where the tree level term $\mathcal{F}_0 \equiv \mathcal{F}_{0,0}$ is the 5d Seiberg-Witten prepotential of the gauge theory. 
Substituting the expansion \eqref{sec2,1.75} in the NY blowup relations \eqref{sec2,1} and taking the NS limit $\epsilon_2 \to 0$ with fixed $\epsilon_1\equiv \epsilon$ the blowup factor \eqref{sec2,1.75} reads
\begin{align}\label{sec2,2}
&\mathcal{B}_{NS}^{(j,k)}(a,\Lm,\epsilon,d,\beta) \equiv 
\lim_{\epsilon_2 \to 0} \mathcal{B}^{(j,k)}(a,\Lm,\epsilon,\epsilon_2,d,\beta)= \nonumber\\ 
&= q^{d/4-1/8 } e^{\alpha-\frac{\beta\gamma(d+k(j-1)/2)}{\epsilon}} \sum_{n\in \mathbb{Z}+\frac{j}{2}} e^{-\frac{n\rho}{\epsilon}} Z_{SD}^{(k)}(a+n \epsilon,\Lm q^{d+k(j-1)/2},\epsilon,\beta)= \nonumber\\
&= q^{d/4-1/8} e^{\alpha-\frac{\beta\gamma (d+k(j-1)/2)}{\epsilon}} Z_D^{(j,k)}(a,\rho,\Lm q^{d+k(j-1)/2},\epsilon,\beta) \ ,
\end{align}
 where $q=e^{\beta \epsilon}$ and $Z^{(k)}_{SD}$ is the Nekrasov partition function in the self-dual background $\epsilon_1=-\epsilon_2=\epsilon$ and we have defined
\begin{equation}\label{sec2,3}
\rho = \epsilon\pdv{a}W(a,\Lm,\epsilon,\beta)\ ,\quad  \alpha = \pdv{\epsilon}W(a,\Lm,\epsilon,\beta) \ , \quad \gamma= \epsilon\Lm\pdv{\Lm}W(a,\Lm,\epsilon,\beta) \ ,
\end{equation}
and $W$ is the 5d twisted superpotential
\begin{equation}\label{sec2,4}
W(a,\Lm,\epsilon,\beta)=\sum_{k\in \{\frac{1}{2}\}\cup\mathbb{Z}_{\geq0}} \mathcal{F}_{0,k}(a,\Lm,\beta) \epsilon^{2k-1} \ .
\end{equation}
The NS blowup factor \eqref{sec2,2} has precisely the structure of the  Kyiv formula for the $q$-Painlevé $\Tau$-function \cite{Bershtein:2018srt} defined in terms of the 5d dual Nekrasov partition function $Z_D^{(j,k)}$. The initial conditions are not generic but, as we explain in the next section, correspond to select a \emph{zero} of the $q$-Painlevé $\Tau$-function in the odd sector, see formula \eqref{sec2,7}. This implies a time reparametrization where the position of the zero is given by the 5d dimensionless instanton counting scale $z$ and the discrete $q$-Painlevé time corresponds to the parameter $d$ which counts the insertions of the codimension 2 observable $I(E)$. The effect of the CS level $k$ on the time variable is to shift the origin of time $d\to d+k(j-1)/2$. Finally, the normalization of the $\Tau$-function is fixed by the factor $e^\alpha$. We have then 
\begin{equation}\label{sec2,5}
\Tau^{(j,k)}(z q^{d+k(j-1)/2})=q^{-d/4+1/8} e^{\frac{\beta\gamma (d+k(j-1)/2)}{\epsilon}}\mathcal{B}_{NS}^{(j,k)}(d)\ .    
\end{equation}

\subsection{Blowup equations and Wilson loop expansion}\label{5dsection2.4}
As previously discussed, the five-dimensional lift of the topological operator/state correspondence implies that the surface observable $I(E)$ can be expressed as an OPE expansion in terms of the circular Wilson loops around the compactified circle dimension
\begin{equation}
    W_R=\Tr_R P
    %\text{-}
    \exp\int_{S^1} (i A+\varphi\dd t)=\Tr_R P
    %\text{-}
    \exp\int_{S^1} \phi \dd t\ ,
\end{equation}
which define a basis for the 5d chiral ring. Therefore, we have the following blowup equations
\begin{equation}
\hat Z^{(j,k)}(a,\Lm,\epsilon_1,\epsilon_2,d,\beta) =\sum_{R}B_{d,R}^{(j,k)}(\Lm,q_1,q_2)\langle W_{R}\rangle_{\mathbb{C}^2\times S^1_\beta}  Z^{(k)}(a,\Lm,\epsilon_1,\epsilon_2,\beta)\ .
\end{equation}
where the sum is over a suitable set of representations of the gauge group $G$. These equations were conjectured in \cite{Wang:2023zcb} from topological string arguments and further explored in \cite{Kim:2021gyj,Kim:2025qaf}. They are a generalization of the 5d Nakajima-Yoshioka blowup equations \cite{Nakajima:2005fg} which correspond to the case $|d|<r$ where $r$ is the rank of the gauge group.

For $SU(2)$ all representations are generated by the tensor products $R_n=2^{\otimes n}$ of the fundamental representation and we will denote with $W_n$ the corresponding Wilson loop. Furthermore, in the NS limit the expectation value of the product of two Wilson loops $W_l, W_m$ factorizes\footnote{This is consistent with our findings coming from the analysis of the expansion of the $q$-Painlevé $\Tau$-function, see for examples \eqref{sec3,4.5}, \eqref{sec3,7.5}.} as stated in \cite{Huang:2022hdo}
\begin{equation}
\ev{W_l W_m}_{NS}=\ev{W_l}_{NS}\ev{W_m}_{NS}\quad \Rightarrow\quad \ev{W_n}_{NS}=\ev{W_1}^n_{NS}=\Uns^n\ .   
\end{equation}
where
\begin{equation}
\Uns\equiv\ev{W_1}_{NS}\ ,   
\end{equation}
is the expectation value of the fundamental Wilson loop in the NS limit. 

Therefore, the blowup factor in the NS limit is an analytic function of $\Uns$ which depends on the value of the parameter $d$ 
\begin{equation}
\mathcal{B}_{NS}^{(j,k)}(a,\Lm,\epsilon,d,\beta)\equiv B^{(j,k)}_{d}(\Uns,z,q)\ ,
\end{equation}
and has the following structure
\begin{equation}\label{sec2,6}
B^{(j,k)}_{d}(\Uns,z,q)\equiv (q^d z)^{\frac{j}{4}} P^{(j,k)}_{d}(\Uns,z,q)=(q^d z)^{\frac{j}{4}}\sum_{n=0}^{n_{\max}^{(j,k)}(d)}P_{d,n}^{(j,k)}(z,q) \Uns^n \ .
\end{equation}
We observe that in principle the blowup factor $B^{(j,k)}_{d}$ is a series in $\Uns$. However, it turns out that for a fixed $d$ only a finite number of representations enter in the rhs of \eqref{sec2,6} and the $P^{(j,k)}_{d}(\Uns,z,q)$ are polynomials\footnote{We isolated a prefactor $(q^d z)^{\frac{j}{4}}$ because in this way $P^{(j,k)}_{d}(\Uns,z,q)$ become polynomials in $\Uns,q,z$, as will follow from our analysis in section \ref{5dsection3}.} in the fundamental Wilson loop $\Uns$ of degree $n_{\max}(d)\sim d^2/4$, as we find for example in \eqref{sec3,4.75}, \eqref{sec3,7.75}. The origin of this truncation may be due to the finiteness of the quantum cohomology ring and technically it appears as a consequence of the bilinear relations. 

The polynomials $P_{d}^{(j,k)}(\Uns,z,q)$ turn out to be $q$-polynomials in $\Uns, z$ with integer coefficients\footnote{For CS level $k=1$ they turn out to be polynomials in $\tilde z =z q^3$.}.
In particular for $d=0,1,-1$ they are given by the NS limit of the standard NY blowup relations \cite{Nakajima:2005fg,Shchechkin:2020ryb}
\begin{align}\label{sec2,7}
 &P^{(0,k)}_{-1}=1 \ ,\quad P^{(0,k)}_{0}=1 \ , \quad P^{(0,k)}_{1}=1 \ , \nonumber\\
 &P^{(1,k)}_{-1}=1 \ ,\quad  P^{(1,k)}_{0}=0 \ , \quad P^{(1,k)}_{1}=-1 \ .
\end{align}
The first non-trivial dependence on the Wilson loop $\Uns$ appears for $d=\pm 2$ and in the NS limit we have the polynomials%({\color{red}check})
\begin{align}\label{sec2,8}
 &P^{(0,0)}_{\pm2} =1-q^{\pm1}z \ , \quad P^{(1,0)}_{\pm 2}=\mp\Uns\ , \\
 &P^{(0,1)}_{\pm2}=1\ , \quad P^{(1,1)}_{2}=-\Uns\ , \quad P^{(1,1)}_{-2}=\Uns + (q^{-1/2} - q^{1/2}) z\ ,
\end{align}
where we observe that we have the following properties\footnote{We checked the formulas for the NS Wilson loop $\Uns$ under the transformation $q\to q^{-1}$ to order five in the instanton expansion.} under the map $d\to -d,\ q\to q^{-1}$
\begin{align}
&\Uns(a,z,q^{-1})= \Uns(a,z,q)\ , &\text{ for }k=0\ , \nonumber\\
&\tilde \Uns(a,z,q)\equiv\Uns(a,z,q^{-1}) = \Uns(a,z,q) + (q^{-1/2} - q^{1/2}) z\ , &\text{ for }k=1\ , \nonumber \\
&B_{-d}^{(j,0)}(\Uns,z,q)=(-1)^jB_d^{(j,0)}(\Uns,z,q)\ , \quad B_{-d}^{(j,1)}(\Uns,z,q)=(-1)^jB_d^{(j,1)}(\tilde\Uns,z,q)\ .
\end{align}
The polynomials \eqref{sec2,8} do not appear in \cite{Nakajima:2005fg}. For $k=0$ they were found in the context of topological string theory \cite{Wang:2023zcb}, and for $k=1$ we checked them up to order three in the instanton expansion. 

In the following section we will use the relation between the NS blowup factor and the $q$-Painlevé $\Tau$-function to compute the OPE coefficients $B^{(j,k)}_{d}(\Uns,z,q)$ recursively.

From the relation \eqref{sec2,2}, this OPE expansion contains the full informations about the self-dual partition function $Z_{SD}^{(k)}$ of the 5d gauge theory and because it is expressed in terms of Wilson loops it can be easily analyzed at any point of the moduli space, including the strongly coupled regime.

As a final comment, let us observe that the scaling of $n_{\max}(d)$ has a possible origin from an effective contact term $T(U)$ which arises when we map the co-dimension two observables $I(E)$, on $\Sigma\times S^1$, in the low-energy effective theory. Indeed, in the 4d case the product of surface observables $O^{(4d)}(\Omega,{\bf F}^2)$ in the UV is mapped to the IR one $\tilde O^{(4d)}(\Omega,{\bf F}^2)$ up to a contact term \cite{Moore:1997pc} $O^{(4d)}(\Omega,{\bf F}^2)O^{(4d)}(\Omega,{\bf F}^2)\to \tilde O^{(4d)}(\Omega,{\bf F}^2)\tilde O^{(4d)}(\Omega,{\bf F}^2)+T_{4d}$ given by the local observable (for the 4d pure $SU(2)$ theory)
\begin{equation}
T_{4d}=\Lambda\partial_{\Lambda}u=(\Lambda\partial_{\Lambda})^2\mathcal{F}_0=\frac{u}{3}+\frac{\pi^2}{12}\frac{E_2(\tau)}{\omega_1^2}\ ,
\end{equation}
where $\mathcal{F}_0$ is the SW prepotential of the 4d theory and $E_2(\tau)$ is the second Eisenstein series. Similarly in the 5d case we have\footnote{Upon exponentiation the Wick contractions introduced by the contact term correspond to a gaussian prefactor in the partition function.} 
\begin{equation}\label{sec2,6.5}
\ev{I(E)^d}_{\hat{\mathbb{C}}^2\times S^1,NS}^{UV}=e^{\frac{1}{2}\beta^2d^2 T} \ev{\tilde I(E)^d}_{\hat{\mathbb{C}}^2\times S^1,NS}^{IR} \ . 
\end{equation}
The CS level $k$ introduces a further source in the contact term which gets shifted as $d^2 T \to d(d+k(j-1))T$. Assuming that the Wilson loop expansion is governed by the UV contact term contribution $T(\Uns)\sim \log \Uns$ it therefore follows that the blowup factor scales like $B_d\sim \Uns^{d^2}$. In the next section we study the NS blowup factor from the $q$-Painlevé equations and its SW limit and we will check directly the above statement\footnote{From a UV perspective, in the four-dimensional case the truncation of the OPE in terms of the chiral ring generator $\Tr \phi^2$ follows from the counting of zero modes introduced by the surface observable wrapped on the exceptional divisor $E$ and is related to the $U(1)$ $R$-charge anomaly. We expect that the truncation of the Wilson loop expansion could be derived via a $K$-theoretic version of the previous analysis studying the theory on the codimension 2 defect $E\times S^1$.}.

\section{Hurwitz expansions of $q$-Painlev\'e $\Tau$-functions}\label{5dsection3}

As we observed in the previous section the NS blowup factor is related to the $q$-Painlevé $\Tau$-function through \eqref{sec2,5} and from \eqref{sec2,7} we have that it vanishes for $d=0$ in the odd sector $j=1$. This is just the five-dimensional analogue of what already observed in \cite{Bonelli:2024wha} where the chiral ring expansion of the $4d$ NS blowup factor was interpreted as the expansion of the Painlevé $\Tau$-function around a zero which allows to compute the OPE coefficients recursively from the Painlevé Hirota equation. In the same way, the 5d blowup factor coefficients $B^{(j,k)}_{d}$  can be determined by studying the expansion of the $q$-Painlevé $\Tau$-functions $\Tau^{(j,k)}(z)$ around a zero $z_0$ of the odd sector $\Tau$-function $\Tau^{(1,k)}(z)$. 

The $q$-Painlevé equation is a $q$-difference equation for $\Tau^{(j,k)}(z)$ where $z$ is the time variable and corresponds to the value of the instanton counting scale $z$ which follows a discrete flow $z \to z q $. Once we expand around a zero $z_0$ we can reparametrize the time variable as
\begin{equation}
 z=z_0 q^d  \ ,
\end{equation}
From the gauge theory point of view the above shift corresponds precisely to the one introduced by the surface observable $I(E)$ as in \eqref{sec2,1}. The $\Tau$-function, being the NS blowup factor $B^{(j,k)}_{d}$, becomes then a function of the discrete time $d$ and using \eqref{sec2,5} it is possible to rewrite the $q$-difference equation as a recurrence relation for the NS blowup factor coefficients
\begin{equation}
B_d^{(j,k)}=F(B_{d'}^{(j',k)})\ ,\quad d'<d\ , j'=0,1\ ,  
\end{equation}
whose explicit form depends on the specific $q$-Painlevé equation. For the values $d=0,\pm1,\pm2$ the above recurrence is not defined\footnote{This is true for the odd sector $j=1$. The other sector is then completely determined.}. The corresponding undetermined coefficients $B^{(j,k)}_{d}$  are \emph{resonances} that parametrize the initial conditions of the the solution and will be fixed using the relations \eqref{sec2,7}, \eqref{sec2,8}.

As discussed in the previous section the structure of the solution will be given by polynomials in $\Uns$ \eqref{sec2,6}. A non-trivial result which we just checked numerically is that these polynomials have coefficients in the ring $\mathbb{Z}[q]$. From the gauge theory point of view this should correspond to the fact that the BPS states introduced by the defect in the exceptional divisor can be reorganized in terms of Wilson loop BPS states and $B_{n,d}^{(j,k)}$ count the degeneracy of these states for some fixed representation $2^{\otimes n}$ and number of insertions $d$. In the following we will study in detail the equation
 \begin{equation}\label{sec3,1}
\Tau^{(j,k)}(qz)\Tau^{(j,k)}(q^{-1}z) = \Tau^{(j,k)}(z)^2 - z^{1/2}\Tau^{(j+1,k)}\left(q^{k/2} z\right)\Tau^{(j-1,k)}\left(q^{-k/2} z\right)\ , \ j \in\mathbb{Z}_2\ ,
 \end{equation}
associated to 5d $\mathcal{N}=1$ $SU(2)$ pure gauge theory with CS level $k=0,1$ and corresponding to $q$-PIII$_3$ and $q$-PI equation respectively \cite{Bershtein:2018srt}. These equations come from the cluster algebra associated to the mutations of the 5d BPS quiver \cite{Bershtein:2018srt,Bershtein:2017swf,Bonelli:2020dcp}. This quiver can be also constructed starting from the Newton polygon associated to the mirror geometry.

 \subsection{$q$-PIII$_3$ alias $N_f=0, k=0$}
 We start from the $q$-PIII$_3$ equation which corresponds to 5d pure gauge theory with CS level $k=0$. The Hirota equation \eqref{sec3,1} becomes
 \begin{equation}\label{sec3,2}
\Tau^{(j,0)}(qz)\Tau^{(j,0)}(q^{-1}z) = \Tau^{(j,0)}(z)^2 -z^{1/2}\Tau^{(j+1,0)}\left( z\right)\Tau^{(j-1,0)}\left(z\right)\ .
 \end{equation}
 We expand now around a zero defining the map given by \eqref{sec2,5}\footnote{In \eqref{sec2,5} the blowup factor differs from the $\Tau$-function by a ``gauge'' prefactor $\exp(a d)$. However, the equation \eqref{sec3,1} is invariant under these gauge transformations and we can simply redefine the $\Tau$-function reabsorbing the gauge prefactor.}
 \begin{equation}
  z\to z q^d \ , \quad B^{(j,0)}_d=\Tau^{(j,0)}(z q^d)\ .  
 \end{equation}
where we redefined the time variable so that $z$ now parametrizes the position of the zero, and we can rewrite the equation \eqref{sec3,2} as
 \begin{equation}\label{sec3,3}
B_{d+1}^{(j,0)}B_{d-1}^{(j,0)} = B_d^{(j,0)}B_d^{(j,0)}  - z^{1/2}q^{d/2}B_d^{(j+1,0)}B_d^{(j-1,0)}\ .
 \end{equation}
This gives the bilateral recurrence relation
 \begin{align}\label{sec3,4}
B_{d+1}^{(j,0)} =\frac{1}{B_{d-1}^{(j,0)}}\left( B_d^{(j,0)}B_d^{(j,0)}  - z^{1/2}q^{d/2}B_d^{(j+1,0)}B_d^{(j-1,0)}\right)\ , \quad d\geq0\ , \\
B_{d-1}^{(j,0)} =\frac{1}{B_{d+1}^{(j,0)}}\left( B_d^{(j,0)}B_d^{(j,0)}  - z^{1/2}q^{d/2}B_d^{(j+1,0)}B_d^{(j-1,0)}\right)\ , \quad d\leq0\ .
 \end{align}
where the initial conditions are given by the NY blowup equations \eqref{sec2,7} and in particular $B_0^{(1,0)}=0$, as we expand around a zero. This is valid only if $B_{d\mp1}^{(j,0)}\neq 0$ which is true for $d\neq\pm1$ or $j\neq1$. Therefore, the relations \eqref{sec3,4} for $j=1, d=\pm1$ do not determine the resonant coefficients $B_{\pm 2}^{(1,0)}$.
%but give some relation between $B_{\pm 1}^{(1,0)}$ and $B_{\pm 1}^{(0,0)}$ which determine one sector in terms of the other, therefore we can choose e.g. $B_{\pm 1}^{(0,0)}$ as independent coefficients and $B_{\pm 1}^{(1,0)}$ are determined and it is easy to check that they are given by \eqref{sec2,7}. Similarly, setting $d=0$ we have a relation between $B_1^{(j,0)}, B_{-1}^{(j,0)}$.  
As it is clear from \eqref{sec2,8} the resonances $B_{\pm 2}^{(1,0)}$ carry the dependence of the blowup factor on the Wilson loop $\Uns$. Together, the knowledge of the coefficients $B_d^{(j,0)}$ for $d=0,\pm 1, \pm 2$ fix then completely the solution\footnote{These coefficients are not all independent but are sufficient to fix the solution uniquely.}. We observe that the recurrence \eqref{sec3,4} contains some denominators, but when we compute the solution they all cancel and the coefficients are polynomials. This non-trivial cancellation is due to the Laurent phenomenon of the cluster algebra in which the new cluster variables are Laurent polynomials of the original ones \cite{fomin2001laurentphenomenon,Fomin:2016caz}.

Applying the above recurrence, we can compute all the coefficients $B_{d}^{(j,k)}$ which have the following form
\begin{align}\label{sec3,4.5}
&B^{(j,0)}_{d}(\Uns,z,q)=(q^d z)^{\frac{j}{4}}P^{(j,0)}_{d}(\Uns,z,q)=(q^d z)^{\frac{j}{4}}\sum^{n^{(j,0)}_{\max}(d)}_{n=0}P_{d,n}^{(j,0)}(z,q) \Uns^n \ , \\
&B_{-d}^{(j,0)}(\Uns,z,q)=(-1)^j B_d^{(j,0)}(\Uns,z,q^{-1})\ ,
\end{align}
where $P_{d}^{(j,k)}(\Uns,z,q)$ turn out to be polynomials in $z, \Uns,q$ with integer coefficients\footnote{We checked this numerically to high order in $d$ but we don't have a proof.}.
We observed numerically that the maximum power $n_{\max}^{(j,0)}(d)$ of $\Uns$ which appears in the polynomial $P_{d}^{(j,k)}$ is
\begin{equation}\label{sec3,4.75}
n_{\max}^{(j,0)}(d) = \left\lfloor \frac{d^2}{4} \right\rfloor-\chi_4(d+2-2j)\ .
\end{equation}
where 
\begin{equation}\label{sec3,chi}
\chi_p(d)=1 \text{ if } d=0\,{\rm mod}\,\,p\ (d\neq0)\text{ and } \chi_p(d)=0 \text{ otherwise}\ .    
\end{equation}
Furthermore, we have the following selection rule
\begin{equation}\label{sec3,5}
P_{d,n}^{(j,0)}(z,q)=0 \ , \quad \text{ if $n=1-j(d+1)\mod 2$}\ .
\end{equation}
For illustration, the polynomials for $0\leq d \leq 6$ are reported in appendix \eqref{appA.1} and they are in agreement with the topological string result (110) in \cite{Wang:2023zcb} in the NS limit $q_2 \to 1, q_1=q$. 

Finally, we observe that the solution is particularly simple for $\Uns=0$ for which we have the closed formula
\begin{equation}\label{new?}
 B_d^{(j,0)}(U=0,z,q)=(-1)^{j(\lfloor d/2 \rfloor - 1)}(z q^d)^{j/4}\frac{1+(-1)^{j(d+1)}}{2}\prod_{l=0}^{\lfloor\frac{d}{2}\rfloor-1}(1-q^{2l+1}z)^{d-1-2l}\ . 
\end{equation}
 This should be related to the algebraic solution of $q$-PIII$_3$ studied in \cite{DelMonte:2023vwv, Bridgeland:2024saj}. 
 
 \subsection{$q$-PI alias $N_f=0, k=1$}\label{subqPI}
 We consider now $q$-PI equation which is associated to 5d pure gauge theory with CS level $k=1$. The Hirota equation \eqref{sec3,1} reduces to
 \begin{equation}\label{sec3,6}
\Tau^{(j,1)}(qz)\Tau^{(j,1)}(q^{-1}z) = \Tau^{(j,1)}(z)^2 - z^{1/2}\Tau^{(j+1,1)}\left(q^{1/2} z\right)\Tau^{(j-1,1)}\left(q^{-1/2} z\right)\ , \ j \in\mathbb{Z}_2\ .
 \end{equation}
From \eqref{sec2,5} we have the following map
  \begin{equation}
  z\to z q^d \ , \quad B^{(j,1)}_d = \Tau^{(j,1)}(z q^{d+(j-1)/2})\ ,
 \end{equation}
and evaluating \eqref{sec3,5} at $z q^{(j-1)/2}$ we obtain
\begin{equation}\label{sec3,7}
B_{d+1}^{(j,1)}B_{d-1}^{(j,1)} = B_d^{(j,1)}B_d^{(j,1)}  - z^{1/2}q^{d/2+(j - 1)/4} B_d^{(j+1,1)}B_{d-(-1)^j}^{(j-1,1)}\ .
 \end{equation}
From \eqref{sec3,7} we get the following recurrence relations\footnote{With our conventions there is some asymmetry between $j=0,1$. This will be convenient in the following, when we will take the strong coupling $4d$ limit.}
  \begin{align}\label{sec3,7.25}
&B_{d+1}^{(0,1)} = \frac{1}{B_{d-1}^{(0,1)}}(B_d^{(0,1)}B_d^{(0,1)}  - z^{1/2}q^{d/2 - 1/4} B_d^{(1,1)}B_{d-1}^{(1,1)})\ , \\
&B_{d+1}^{(1,1)} = \frac{1}{B_{d-1}^{(1,1)}}(B_d^{(1,1)}B_d^{(1,1)}  - z^{1/2}q^{d/2} B_d^{(0,1)}B_{d+1}^{(0,1)})\ .
 \end{align}
And again the coefficients $B_{\pm 2}^{(1,1)}$ are resonances and we need their value to fix completely the solution. These are again fixed by \eqref{sec2,8}. The solution has the following structure 
\begin{align}\label{sec3,7.5}
&B^{(j,1)}_{d}(\Uns,z,q)=(z q^d)^{j/4}P^{(j,1)}_{d}(\Uns,z,q)=(z q^d)^{j/4}\sum^{n_{\max}^{(j,1)}(d)}_{n=0}P_{d,n}^{(j,1)}(zq^{3/2},q) \Uns^n \ , \\
&B_{-d}^{(j,1)}(\Uns,z,q)=(-1)^j B_d^{(j,1)}(\tilde \Uns,z,q^{-1})\ ,  \quad \tilde\Uns=\Uns + (q^{-1/2} - q^{1/2}) z\ ,
\end{align}
and again $P^{(j,1)}_{d}(\Uns,z,q)$ turn out to be integer $q$-polynomials in $\Uns,\tilde z=z q^{3/2}$ and numerically the maximum power $n_{\max}^{(j,1)}(d)$ of $\Uns$ turns out to be
\begin{equation}\label{sec3,7.75}
n_{\max}^{(j,1)}(d) = \left\lfloor \frac{d(d+j-1)}{4}\right\rfloor -j\chi_4(d)\ .
\end{equation}
In this case there is no selection rule like \eqref{sec3,5}. The first polynomials $P_{d}^{(j,1)}$ are reported in appendix \eqref{appA.2}.

\subsection{Blowup factor in the SW limit}\label{SW}
To understand better some features of the Wilson loop Hurwitz expansion, let us now study the SW limit $\epsilon\to 0$ of the NS blowup factor \eqref{sec2,2}. Substituting the genus expansion \eqref{sec2,1.75} in \eqref{sec2,2} and following the same steps of \cite{Bonelli:2024wha} we arrive at the following result
\begin{align}\label{sec2,2.5}
&\mathcal{B}_{SW}^{(j,k)}(a,\Lm,d,\beta)=\lim_{\epsilon\to 0}\mathcal{B}_{NS}^{(j,k)}(a,\Lm,\epsilon,d,\beta)=\\
&=-e^{\mathcal{F}_{1,0}+\mathcal{F}_{0,1}}\exp\left[\frac{1}{2}\beta^2(d+k(j-1)/2)^2 T\right]\theta_{4-3j}\left( h\beta(d+k(j-1)/2)\eval\tau\right)\ , \nonumber
\end{align}
where 
\begin{equation}
\tau(a,\Lm,\beta) =\frac{1}{2\pi i}\pdv[2]{\mathcal{F}_0}{a}\left(a,\Lm,\beta\right)\ , \quad h(a,\Lm,\beta)=\frac{1}{2i}\Lm \pdv[2]{\mathcal{F}_0}{\Lm}{a}  \ , \quad T(a,z,\beta)=(\Lm\partial_{\Lm})^2\mathcal{F}_0(a,z,\beta) \ .
\end{equation}
and we identify $T$ with the contact term introduced by the 5d codimension 2 observable $I(E)$ that we discussed in subsection \ref{5dsection2.4}.
is the IR coupling and we used the Jacobi theta functions 
\begin{equation}\label{2.2}
\theta_1(x|\tau)= -\sum_{n\in \mathbb{Z}+\frac{1}{2}} (-1)^{n} e^{i\pi\tau n^2+2nix} \ , \quad \theta_4(x|\tau)= \sum_{n\in \mathbb{Z}} (-1)^{n} e^{i\pi \tau n^2+2nix}\ .
\end{equation}
    
We can now check the contact term scaling behaviour discussed in the end of the previous section. For simplicity we consider the case $k=0$. To do this we compare the explicit expression of the blowup factor \eqref{sec2,2.5} with its expression in terms of polynomials in the SW Wilson loop vev $U$ 
\begin{equation}
B^{(j,k)}_{d,SW}(U,z)=\lim_{q\to1}B^{(j,k)}_{d}(\Uns,z,q)\ , \quad \mathcal{B}_{SW}^{(j,k)}(a,z,d,\beta)=B^{(j,k)}_{d,SW}(U,z)
\end{equation}
In particular from \eqref{sec2,6}, \eqref{sec2,7}, \eqref{sec2,8} in the limit $q\to1$ for $k=0$ we have the following equations
\begin{equation}
B_{0,SW}^{(0,0)}(U,z)=1\ ,\quad B_{1,SW}^{(0,0)}(U,z)=1 \ , \quad B_{1,SW}^{(1,0)}(U,z)=-z^{1/4} \ , \quad B_{2,SW}^{(1,0)}(U,z)=-z^{1/4}U\ , 
\end{equation}
and substituting the explicit expression \eqref{sec2,2.5} we get
\begin{align}\label{sec3,10.25}
&d=0:\quad e^{\mathcal{F}_{1,0}+\mathcal{F}_{0,1}}=-\frac{1}{\theta_4(0|\tau)}\ , \\
&d=1:\quad e^{-\frac{1}{2}\beta^2 T}=\frac{\theta_4\left(\beta h\eval\tau\right)}{\theta_4\left(0\eval\tau\right)}=-z^{-1/4}\frac{\theta_1\left(\beta h\eval\tau\right)}{\theta_4\left(0\eval\tau\right)}\ , \\
&d=2: \quad U=-z^{-1/4}e^{2\beta^2T}\frac{\theta_1(2\beta h|\tau)}{\theta_4(0|\tau)}\ ,
\end{align}
which implies
\begin{equation}
z^{1/4}=-\frac{\theta_1\left(\beta h\eval\tau\right)}{\theta_4\left(\beta h\eval\tau\right)}\ , \quad U=\frac{\theta_4^{3}\left(0\eval\tau\right)}{\theta_4^{3}\left(\beta h\eval\tau\right)}\frac{\theta_1(2\beta h|\tau)}{\theta_1\left(\beta h|\tau\right)} \ , \quad e^{-\frac{1}{2}\beta^2 T}=\frac{\theta_4\left(\beta h\eval\tau\right)}{\theta_4\left(0\eval\tau\right)}\ ,
\end{equation}
and
\begin{equation}
\quad e^{\frac{1}{2}\beta^2T}=U^{1/4} \left(\frac{\theta_1(2\beta h|\tau)}{\theta_1\left(\beta h\eval\tau\right)} \frac{\theta_4\left(\beta h\eval\tau\right)}{\theta_4(0|\tau)}\right)^{-1/4}       
\end{equation}
which is the five-dimensional analogue of the 4d contact term equation
\begin{equation}
T_{4d}=\Lambda\partial_{\Lambda}u=(\Lambda\partial_{\Lambda})^2\mathcal{F}_0=\frac{u}{3}+\frac{\pi^2}{12}\frac{E_2(\tau)}{\omega_1^2}\ ,
\end{equation}
and reduces to it in the limit $\beta\to0$ (at second order in $\beta$). Using the above relations the blowup factor for $j=1,k=0$ reads
\begin{equation}\label{sec3,10.5}
\mathcal{B}_{SW}^{(j,0)}(a,\Lm,d,\beta)=B^{(j,0)}_{d,SW}(U,z)=U^{d^2/4} \left(\frac{\theta_1(2\beta h|\tau)}{\theta_1\left(\beta h\eval\tau\right)} \frac{\theta_4\left(\beta h\eval\tau\right)}{\theta_4(0|\tau)}\right)^{-d^2/4}      \frac{\theta_{4-3j}\left(\beta d h\eval\tau\right)}{\theta_4(0|\tau)}\ ,
\end{equation}
this is manifestly modular invariant if rewritten in terms of Weierstrass $\sigma$-function as
\begin{equation}\label{sec3,10.5bis}
\mathcal{B}_{SW}^{(j,0)}(a,\Lm,d,\beta)=B^{(j,0)}_{d,SW}(U,z)=U^{d^2/4} \left(\frac{\sigma(2\beta)}{\sigma\left(\beta\right)} \frac{\sigma\left(\beta+\omega_2\right)}{\sigma(\omega_2)}\right)^{-d^2/4}      \frac{\sigma\left(\beta d+\omega_2(1-j)\right)}{\sigma(\omega_2)}\ ,
\end{equation}
where the Weierstrass $\sigma$-function is
\begin{equation}\label{sigma}
\sigma(s;\omega_1,\omega_2) = \frac{2\omega_1}{\pi} e^{\frac{\pi^2}{6}\frac{E_2(\tau)}{4\omega_1^2}s^2}\frac{\theta_1\left(\frac{\pi s}{2\omega_1}\right)}{\theta_1'(0)} \ , \quad \theta_1'(0)=2\eta(\tau)^3 \ , \quad \omega_1=\frac{1}{2h}\ , \quad \omega_2=\frac{\tau}{2h} \ .
\end{equation}
which suggests a quadratic scaling $\mathcal{B}_{SW}^{(j,0)}(a,\Lm,d,\beta)\sim U^{d^2/4}$ as previously discussed\footnote{A similar analysis can be done for the case $k=1$. The results are in agreement with the autonomous limit $q\to1$ of $q$-PI Hurwitz expansion that we found in \ref{subqPI}.} and from \eqref{sec3,10.25} we see that this is given precisely by the contribution of the contact term.

We observe that the actual scaling cannot be $d^2/4$ because in general it is rational, which violates holomorphicity in $U$, and in principle it can be modified by the contributions coming from the $\theta$-functions in \eqref{sec3,10.5}. This is indeed the case, and the effect of this is precisely to make the maximal power of $U$ integer with the addition of a suitable function of $d$ which depends on the sector and does not change the qualitative quadratic growing.

To verify this statement we can estimate the maximal power of \(U\) appearing in \(B_{d,SW}^{(j,0)}\) by analyzing the following limit of the SW parameters
\begin{equation}\label{scalim}
\beta h=\frac{\tau}{4}+x,\qquad \tau\to i\infty\ ,
\end{equation}
which gives the following asymptotics of the theta functions
\begin{equation}\begin{gathered}
\theta_1(\beta h|\tau)=i e^{-\pi x}+O(e^{\frac{\pi i \tau}{2}}), \qquad \theta_4(\beta h|\tau)=1+O(e^{\frac{\pi i \tau}{2}}),\\ \theta_1(2\beta h|\tau)=e^{2\pi x}e^{-\frac{\pi i\tau}{4}}+O(e^{\frac{3\pi i\tau}{4}}),\qquad \theta_4(0|\tau)=1+O(e^{\pi i \tau})\ .
\end{gathered}\end{equation}
In this limit the parameters \(U\) and \(z\) have the following asymptotics
\begin{equation}
z^{1/4}\sim -i e^{-\pi x}, \qquad U\sim -i e^{-\frac{\pi i\tau}{4}}e^{\pi x}\ ,
\end{equation}
which means that the limit \eqref{scalim} corresponds to \(U\to\infty\) with \(z\) fixed and therefore in this limit we can directly see the polynomial degree in \(U\).

The limit of the blow-up factor can be done with a saddle-point approximation of the $\theta$ series \eqref{2.2} and it gives
\begin{equation}
B_{d,SW}^{(j,0)}(U,z)=\frac{\theta_4(0|\tau)^{d^2}}{\theta_4(\beta h|\tau)^{d^2}}\frac{\theta_1(\beta dh|\tau)}{\theta_4(0|\tau)}\sim \theta_{4-3j}(\beta dh|\tau)\sim e^{i \pi\tau f_j(d)}.
\end{equation}
where $f_j(d)$ is the value at the saddle-point
\begin{equation}
f_j(d)=\min\limits_{n\in \mathbb{Z}}\left((n+d/4+j/2)^2-d^2/16 \right)\ .  
\end{equation}
From the previous asymptotic analysis we obtain that the leading order of $U$ in \(B_{d,SW}^{(j,0)}(U,z)\) is \(U^{n_{\max}^{(j,0)}(d)}\) with
\begin{equation}
n_{\max}^{(j,0)}(d)=\frac{d^2}{4}-\min\limits_{n\in \mathbb{Z}}(2n+d/2+j)^2 = \left\lfloor\frac{d^2}{4}\right\rfloor-\chi_4(d+2-2j)\ ,
\end{equation}
which is in agreement with the $q$-PIII$_3$ result \eqref{sec3,4.75}, in particular, the maximal power of $U$ is not modified by the $\Omega$-background corrections.

\paragraph{Expansion around the 5d superconformal point}

The SW blowup factor \eqref{sec3,10.5} has an interesting structure in the strongly coupled superconformal point $z=1$. In this case we have
\begin{equation}\label{sec3,10.625}
B^{(j,0)}_{d}=(-1)^{\left\lfloor \frac{d}{4} \right\rfloor+j}U^{\left\lfloor \frac{d^2}{4} \right\rfloor}(1-\chi_4(d+2-2j))\ ,
\end{equation}
which gives the scaling $n_{\max}^{(j,0)}(d) = \left\lfloor \frac{d^2}{4} \right\rfloor-\chi_4(d+2-2j)$ as observed in the $q$-Painlevé case \eqref{sec3,4.75}.

It is natural to ask if there are other points in the moduli space where the blowup factor has this simple structure. In the following we will find these special points studying the SW blowup factor from the autonomous limit $q\to1$ of the $q$-Painlevé equations\footnote{In principle one can consider also the case with $\Omega$-background $q\neq1$ but we didn't find any simple structure in this case.}.

\subsubsection{Special points for $k=0$}
 For local $\mathbb{F}_0$ ($k=0$) the SW curve is
\begin{equation}
 e^{p}+e^{-p}+e^x+ze^{-x}+U=0  \ .
\end{equation}
The corresponding $q$-PIII$_3$ equation \eqref{sec3,3} in the autonomous, i.e. SW, limit $q\to1$ is
 \begin{equation}\label{sec3,10.75}
B_{d+1,SW}^{(j,0)}B_{d-1,SW}^{(j,0)} = B_{d,SW}^{(j,0)}B_{d,SW}^{(j,0)}  - z^{1/2}B_{d,SW}^{(j+1,0)}B_{d,SW}^{(j-1,0)}\ .
 \end{equation}
To find the special points we consider the ansatz
\begin{equation}\label{sec3,sp-1}
 B_{d,SW}^{(j,0)}=z^{j/4}b_d^{(j)}U^{\left\lfloor \frac{d^2}{p} \right\rfloor}\ , \quad b_d^{(j)}=0,\pm1\ ,
\end{equation}
where $p$ is a positive integer and $b_d^{(j)}$ is $p$-periodic/anti-periodic
\begin{equation}
b_{d+p}^{(j)}=s_jb_d^{(j)}\ ,\quad s_j=\pm 1 \ .
\end{equation}
The NY blowup equations \eqref{sec2,7} fix the initial conditions
\begin{align}\label{sec3,sp0}
 &b^{(0,0)}_{-1}=1 \ ,\quad b^{(0,0)}_{0}=1 \ , \quad b^{(0,0)}_{1}=1 \ , \nonumber\\
 &b^{(1,0)}_{-1}=1 \ ,\quad b^{(1,0)}_{0}=0 \ , \quad b^{(1,0)}_{1}=-1\ ,
\end{align}
and from \eqref{sec2,8} we have the condition
\begin{equation}\label{sec3,sp1}
B^{(1,0)}_{2}=-z^{1/4}U=z^{1/4}b_2^{(1,1)}U^{\left\lfloor \frac{4}{p} \right\rfloor}\quad \Rightarrow\quad b_2^{(1,1)}=-1 \ , \quad\left\lfloor \frac{4}{p}\right\rfloor=1\quad \Rightarrow \quad p=3,4\ .
\end{equation}
Substituting the ansatz \eqref{sec3,sp-1} in \eqref{sec3,10.75} we get
\begin{align}
&b_{d+1}^{(0,0)}b_{d-1}^{(0,0)} U^{\Delta_p(d)}=b_d^{(0,0)}b_d^{(0,0)}  - z b_d^{(1,0)}b_d^{(1,0)}\ , \label{sec3,sp2,0}\\
&b_{d+1}^{(1,0)}b_{d-1}^{(1,0)} U^{\Delta_p(d)}=b_d^{(1,0)}b_d^{(1,0)} -b_d^{(0,0)}b_d^{(0,0)}\ . \label{sec3,sp2,1}
\end{align}
where 
\begin{equation}
 \Delta_p(d)={\left\lfloor \frac{(d+1)^2}{p} \right\rfloor}+{\left\lfloor \frac{(d-1)^2}{p} \right\rfloor}-2{\left\lfloor \frac{d^2}{p} \right\rfloor}\ , 
\end{equation}
which is $p$-periodic and for $p\geq 3$ takes values in $\{0,1,-1\}$
\begin{equation}
\Delta_p(d+p)=\Delta_p(d)\ , \quad \Delta_p(d)=0,\pm1 \text{ for }p\geq 3 
\end{equation}
 From \eqref{sec3,sp0} and the equation \eqref{sec3,sp2,0} for $d=1$ we get the condition
\begin{equation}
b_2^{(0,0)} U^{\Delta_p(1)}=1-z \quad \Rightarrow \quad z=1- b_2^{(0,0)} U^{\Delta_p(1)}\ .  
\end{equation}
We can choose $b_2^{(0,0)}=0,\pm1$ and we have $\Delta_p(1)=1$ for $p=3,4$ so the only possible special points are
\begin{equation}
z=1\ , \quad z=1\pm U \ ,  
\end{equation}
which can have $\mathbb{Z}_p$ symmetry with $p=3,4$. To fix the period we have to verify if we get $b_p^{(j,0)}=s_jb_0^{(j,0)},\ b_{p-1}^{(j,0)}=s_jb_{-1}^{(j,0)}$ for some $p$\footnote{Indeed, from \eqref{sec3,sp2,0}, \eqref{sec3,sp2,1}, we have an autonomous recurrence and $b_{d+1}^{(j,0)}$ has the same sign of $b_{d-1}^{(j,0)}$. Therefore, $b_d^{(j,0)}$ will be periodic/anti-periodic if at some point we go back to the initial values up to the sign $s_j$. We also observe for the point $z=1$ we have $b_2^{(0,0)}=0$ and the recurrence relation is ill-defined and should be regularized. This is done setting $b_2^{(0,0)}=\delta$ and taking the limit $\delta\to0$ at the end.}. By a direct check we find that $z=1$ correspond to the $\mathbb{Z}_4$ symmetric solution \eqref{sec3,10.625}
\begin{equation}
B^{(j,0)}_{d}=(-1)^{\left\lfloor \frac{d}{4} \right\rfloor+j}U^{\left\lfloor \frac{d^2}{4} \right\rfloor}(1-\chi_4(d+2-2j))\ ,
\end{equation}
therefore, we recover the superconformal point that we previously discussed. In the same way, we find that the two points $z=1\pm U$ correspond to the $\mathbb{Z}_3$ symmetric solutions\footnote{It seems that some structure survives if we choose $q^m=1$ for some integer $m$. For instance for $q^4=1$ some of the zeros survive and the coefficients have a simple structure $B_d\sim U^n ((-4-4 i)+U^2)^l)$.}
\begin{equation}
B^{(0,0)}_{d}=U^{\left\lfloor \frac{d^2}{3} \right\rfloor}\ ,\quad B^{(1,0)}_{d}=-(1\pm U)^{1/4}(-1)^{\left\lfloor \frac{d}{3} \right\rfloor}U^{\left\lfloor \frac{d^2}{3}\right\rfloor}(1-\chi_3(d))\ .
\end{equation}

\subsubsection{Special points for $k=1$}
For local $\mathbb{F}_1$ ($k=1$) the SW curve is
\begin{equation}
 e^{p}+e^{-p+x}+e^x+ze^{-x}+U=0  \ , 
\end{equation}
and the corresponding $q$-PI equation \eqref{sec3,7} in the autonomous limit $q\to1$ reads
\begin{equation}\label{sec3,10.875}
B_{d+1,SW}^{(j,1)}B_{d-1,SW}^{(j,1)} = B_{d,SW}^{(j,1)}B_{d,SW}^{(j,1)}  - z^{1/2} B_{d,SW}^{(j+1,1)}B_{d-(-1)^j,SW}^{(j-1,1)}\ .
 \end{equation}
In this case we consider the ansatz
\begin{align}
 &B_d^{(0,1)}=b_d^{(0,1)}U^{\left\lfloor \frac{d(d-1)}{p} \right\rfloor}\ , \quad b_d^{(0,1)}=0,\pm1\ , \\
 &B_d^{(1,1)}=(q^dz)^{1/4}b_d^{(1,1)}U^{\text{nint}({\frac{d^2}{p}})}\ , \quad b_d^{(1,1)}=0,\pm1\ , 
\end{align}
where nint$(x)$ is the nearest integer function\footnote{For $p=3,4$ we have nint$(\frac{d^2}{p})=\left\lfloor \frac{d^2}{p} \right\rfloor$.}and the initial conditions \eqref{sec2,7} 
\begin{align}%\label{sec3,sp0}
 &b^{(0,1)}_{-1}=1 \ ,\quad b^{(0,1)}_{0}=1 \ , \quad b^{(0,1)}_{1}=1 \ , \nonumber\\
 &b^{(1,1)}_{-1}=1 \ ,\quad b^{(1,1)}_{0}=0 \ , \quad b^{(1,1)}_{1}=-1\ ,
\end{align}
and \eqref{sec2,8}
\begin{equation}%\label{sec3,sp1}
B^{(1,1)}_{2}=-z^{1/4}U=z^{1/4}b_2^{(1,1)}U^{\text{nint}(\frac{4}{p})}\quad \Rightarrow\quad b_2^{(1,1)}=-1 \ , \quad\text{nint}\left(\frac{4}{p}\right)=1\quad \Rightarrow \quad p=3,4,5,6,7\ .
\end{equation}
and the equation \eqref{sec3,10.875} becomes 
\begin{align}
&b_{d+1}^{(0,1)}b_{d-1}^{(0,1)} U^{\Delta'_p(d)}= b_d^{(0,1)}b_d^{(0,1)} - zb_d^{(1,1)}b_{d-1}^{(1,1)}U^{\Delta''_p(d)}\ ,\label{sec3,sp3,0}\\
&b_{d+1}^{(1,1)}b_{d-1}^{(1,1)} U^{{\text{nint}\left(\frac{(d+1)^2}{p} \right)}+{\text{nint}\left(\frac{(d-1)^2}{p} \right)}}= U^{2\text{nint}\left(\frac{d^2}{p} \right)}b_d^{(1,1)}b_d^{(1,1)} -b_d^{(0,1)}b_{d+1}^{(0,1)}U^{\left\lfloor \frac{(d+1)d}{p} \right\rfloor+\left\lfloor \frac{d(d-1)}{p} \right\rfloor}\ , \label{sec3,sp3,1}
\end{align}
where we defined
\begin{align}
&\Delta'_p(d)={\left\lfloor \frac{(d+1)d}{p} \right\rfloor}+{\left\lfloor \frac{(d-1)(d-2)}{p} \right\rfloor}-2{\left\lfloor \frac{d(d-1)}{p} \right\rfloor}\ , \\
&\Delta''_p(d)=\text{nint}\left(\frac{d^2}{p} \right)+\text{nint}\left(\frac{(d-1)^2}{p} \right)-2\left\lfloor \frac{d(d-1)}{p}\right\rfloor\ ,
\end{align}
that are again $p$-periodic. For $d=1$ we have $\Delta'_p(1)=\Delta''_p(1)=0$ for $p\geq 3$ and the equation \eqref{sec3,sp3,0} gives
\begin{equation}
b_{2}^{(0,1)}= 1\ ,
\end{equation}
and for $d=2$ we have the condition
\begin{equation}
b_{3}^{(0,1)}U^{\Delta'_p(2)}= 1- z U^{\Delta''_p(2)}\quad \Rightarrow \quad z=U^{-\Delta''_p(2)}-b_{3}^{(0,1)}U^{\Delta'_p(2)-\Delta''_p(2)}    
\end{equation}
We can choose $b_3^{(0,1)}=0,\pm1$ and we have
\begin{equation}
\Delta_p'(2)=\begin{cases}
2\ \text{ if }p=3  \\
1\ \text{ if }p=4,5,6 \\ 
0\ \text{ if }p>6 
\end{cases}    
\ \ , \quad
\Delta_p''(2)=\begin{cases}
1\ \text{ if }p=3,4,5\\
0\ \text{ if }p>5 \ ,
\end{cases}  
\end{equation}
so the possible special points are
\begin{align}
&z=U^{-1}\ ,\quad  z=U^{-1}\pm U   \quad \text{for }p=3\ , \\
&z=U^{-1}\ , \quad z=U^{-1}\pm 1   \quad \text{for }p=4,5 \ \\
&z=1 \ , \quad z=1\pm U\ ,\quad \text{for }p=6\ , \\
&z=0,1,2 \quad \text{for }p>6 \ .
\end{align}
Again we need to verify if the recurrence closes at some $p$, $b_p^{(j,1)}=s_jb_0^{(j,1)},\ b_{p-1}^{(j,1)}=s_jb_{-1}^{(j,1)}$.
This happens only for the point $z=U^{-1}-U$ which gives the $\mathbb{Z}_3$ symmetric solution
\begin{equation}
B^{(0,1)}_{d}=U^{\left\lfloor \frac{d(d-1)}{3} \right\rfloor}\ ,\quad B^{(1,1)}_{d}=-(U^{-1}-U)^{1/4}(-1)^{\left\lfloor \frac{d}{3} \right\rfloor}U^{\text{nint}\left( \frac{d^2}{3}\right)}(1-\chi_3(d))\ .
\end{equation}
and the point $z=U^{-1}$ which corresponds to the $\mathbb{Z}_5$ symmetric solution
\begin{equation}
B^{(0,1)}_{d}=(-1)^{\left\lfloor \frac{d}{5} \right\rfloor}U^{\left\lfloor \frac{d(d-1)}{5} \right\rfloor}(1-\chi_5(d+3))\ , \quad 
B^{(1,1)}_{d}=-U^{-1/4}(-1)^{\left\lfloor \frac{d}{5} \right\rfloor}U^{\text{nint}\left(\frac{d^2}{5} \right)}(1-\chi_5(d))\ . \quad 
\end{equation}

\section{Limits to four-dimensional theories}\label{5dsection4}
In this section we study the four-dimensional reduction of 5d SYM with CS level $k=0,1$ and there are two possible limits we can consider. 

The UV completions of the 5d $\mathcal{N}=1$ SYM with CS level $k=0$ and $k=1$ are the so-called $E_1$ and $\tilde E_1$ SCFTs \cite{Seiberg:1996bd}. In both cases, this SCFT has a single real mass deformation $M_{UV}=1/g^2_5$, which can be positive or negative, related to the instanton mass i.e. the 5d UV bare coupling. The relation with the 5d instanton counting parameter $z$ is
\begin{equation}
 z= e^{-\frac{8\pi^2\beta}{g_5^2}}\ .  
\end{equation}

The first limit we can take is the standard geometric engineering limit where we integrate out the massive KK modes keeping the combination $\Lambda^4=\beta^{-4}z$ finite, so we are in a weakly coupled phase of the 5d theory. In this way we recover 4d $\mathcal{N}=2\, SU(2)$ pure SYM with coupling $\Lambda$. Notice that this limit is independent on the value of the CS level $k$. 

The other limit we consider is obtained by doing a dimensional reduction $\beta\to 0$ at \emph{finite} 5d coupling $z$. This gives a non trivial result only for $k=1$, as in the $k=0$ theory this limiting procedure leads to a purely topological theory as the RGE become trivial. This is explained in full detail from this view point in subsection \ref{5dsection4.2}.

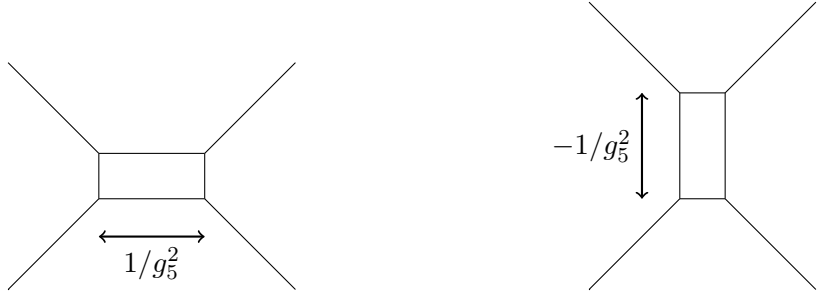
\begin{figure}[ht]
\centering
\begin{tikzpicture}
 \pqwebbody{(-0.2cm,-0.2cm)}{(0.2cm,-0.2cm)}{(0.2cm,0.2cm)}{(-0.2cm,0.2cm)}{(0,0)}{(0,0)}{(0,0)}{(0,0)}{0}
 \draw[<->, thick] ($(C)+(0,-0.5cm)$)--($(D)+(0cm,-0.5cm)$) node[pos=0.5, below] {$1/g_5^2$};
\end{tikzpicture}
\hspace{3cm}
\begin{tikzpicture}
 \pqwebbody{(0.2cm,0.2cm)}{(-0.2cm,0.2cm)}{(-0.2cm,-0.2cm)}{(0.2cm,-0.2cm)}{(0,0)}{(0,0)}{(0,0)}{(0,0)}{0}
 \draw[<->, thick] ($(A)+(-0.5cm,0)$)--($(D)+(-0.5 cm,0)$) node[pos=0.5, left] {$-1/g_5^2$};
\end{tikzpicture}

\caption{\label{fig2:pqweb0} $(p,q)$-web for 5d SYM with $CS$ level $k=0$. For negative coupling the roles of instantons and $W$-bosons are exchanged.}
\end{figure}

Indeed, in the case of $E_1$ SCFT, for $M_{UV}>0$ the theory flows to the 5d pure theory with $k=0$ which is weakly coupled. For a negative mass deformation $M_{UV}<0$ we obtain a physically equivalent theory and again we flow to the weakly coupled 5d $SU(2)$ SYM theory with $k=0$ where instanton particles and W-bosons are just exchanged. The new parameters become
\cite{Aharony:1997bh}
\begin{equation}
\tilde m_I=m_W,\quad \tilde m_W=m_I \iff \tilde M_{UV}=1/\tilde g_5^2=-1/g_5^2=-M_{UV}\ ,\quad \tilde a=a+1/g_5^2\ .
\end{equation}
In the $(p,q)$-web construction, for $M_{UV}<0$ the roles of D5 branes and NS5 branes are inverted, see figure \ref{fig2:pqweb0} where the horizontal lines represent D5 branes and the vertical ones represent NS5 branes. This means that for $k=0$ there is no strongly coupled phase.

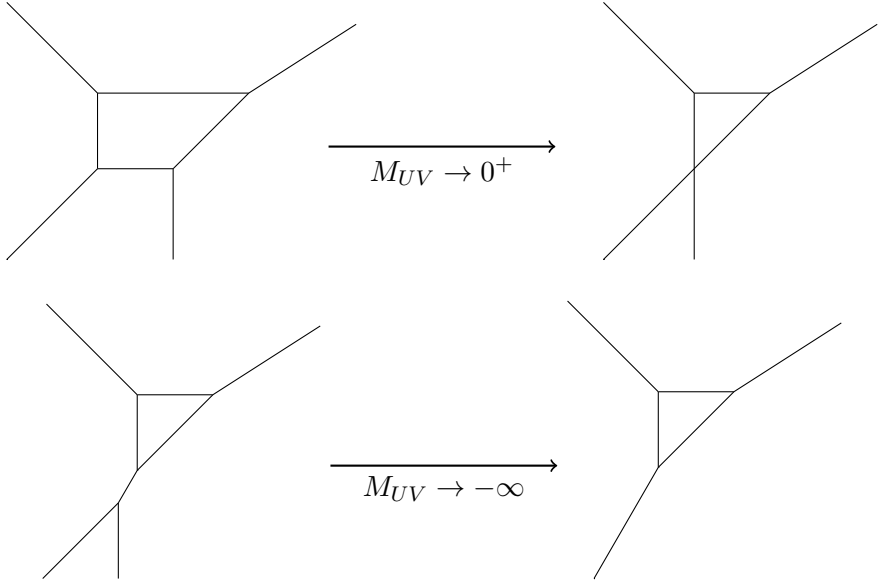
\begin{figure}[h]
\centering
\pqweb{(0,0)}{(1cm,0)}{(0,0)}{(0,0)}{(0,0)}{(0.22cm,-0.29cm)}{(-1.2cm,0)}{(0,0)}{0}  
\begin{tikzpicture}
\useasboundingbox (0,0) rectangle (3cm,3.5cm);
\coordinate (A) at ($(-0.5,1.5)$);
\draw[thick,->] (A)--++(3,0) node[pos=0.5, below] {$M_{UV}\to0^+$}; 
\end{tikzpicture}
\pqweb{(1cm,0)}{(1cm,0)}{(0,0)}{(1cm,0)}{(0,0)}{(0.22cm,-0.29cm)}{(-1.2cm,0)}{(0,0)}{0} \\
\vspace{0.5cm}
\pqweb{(1cm,0)}{(1cm,0)}{(0,0)}{(1cm,0)}{(0,0)}{(0.22cm,-0.29cm)}{(-1.2cm,1.2cm)}{(0.95cm,0.77cm)}{1}
%\hspace{2cm}
\begin{tikzpicture}
\useasboundingbox (0,0) rectangle (3cm,3.5cm);
\coordinate (A) at ($(0,1.5)$);
\draw[thick,->] (A)--++(3,0) node[pos=0.5, below] {$M_{UV}\to-\infty$}; 
\end{tikzpicture}
\pqweb{(1cm,0)}{(1cm,0)}{(0,0)}{(1cm,0)}{(0,0)}{(0.22cm,-0.29cm)}{(-1.2cm,1.2cm)}{(0.35cm,-0.27cm)}{0}
\caption{\label{fig3:pqweb1}$(p,q)$-web for 5d SYM with $CS$ level $k=1$ for $M_{UV}>0$ (upper part) and $M_{UV}<0 $ (lower part). In the limit $M_{UV}\to-\infty$ we obtain the $E_0$ theory (local $\mathbb{P}^2$ geometry).}        
\end{figure}

The situation is drastically different in the case of $\tilde E_1$ SCFT. In this case the flow triggered by the mass deformation depends on the sign of $M_{UV}$. For $M_{UV}>0$ the theory flows to 5d SYM with CS level $k=1$ so we land in a weakly coupled gauge theory phase. For $M_{UV}<0$ instead the theory flows to the non-lagrangian $E_0$ SCFT which has no flavour symmetry and corresponds to 
a deformed version of the local $\mathbb{P}^2$ geometry, obtained from the local $\mathbb{F}_1$ geometry, (see figure \ref{fig3:pqweb1}), the undeformed one being reached in the limit $M_{UV}\to-\infty\iff z\to \infty$. We will see that in this strongly coupled region, at finite coupling $z$, the theory admits a non trivial 4d limit to the AD theory H$_0$.

At the level of the $q$-Painlevé $\Tau$-function this difference arises from an extra symmetry $z \to 1/z$ of the equation \eqref{sec3,1} for $k=0$ which is absent for $k=1$ \cite{Bershtein:2016uov}.

We want now to analyze in detail these limits at the level of the RGEs.

 \subsection{Geometric engineering limit}
 For any $k$ we can recover the standard $4d$ limit to pure $SU(2)$ gauge theory and the corresponding NS blowup factor by the geometric engineering limit corresponding to Kaluza-Klein (KK) reduction on the circle direction. More precisely, in the 4d limit we scale
 \begin{equation}
  \beta=\frac{\log q}{\epsilon}\to 0\iff q\to1\ ,  
 \end{equation}
 keeping fixed the 4d instanton scale $\Lambda^4$
 \begin{equation}
\quad z=\beta^4\Lambda^4=\frac{(\log q)^4}{\epsilon^4}\Lambda^4=\frac{(q-1)^4}{\epsilon^4}\Lambda^4+O((q-1)^5)\ .
 \end{equation}
 This limit corresponds to one in which the 4d UV theory is weakly coupled. Indeed, the UV coupling $g_4(1/\beta)$ of the 4d theory at the scale of the circle radius $\mu=1/\beta$ given by RG matching is
\begin{equation}
\frac{1}{g_4^2(1/\beta)}=\frac{\beta}{g_5^2}\ ,
\end{equation}
and the RG invariant scale $\Lambda$ is then
\begin{equation}
 \Lambda^4=\frac{1}{\beta^4}e^{-\frac{8\pi^2}{g_4^2(1/\beta)}}\ . 
\end{equation}
Therefore, in the limit $\beta\to0$ at fixed $\Lambda$ we have $g_4^2(1/\beta)\to 0$ in agreement with 4d asymptotic freedom. From a 5d point of view in the limit $\beta\to0$ we send the ratio between the UV mass deformation of the 5d SCFT $M_{UV}=1/g_5^2$ and the KK mass scale $M_{KK}=1/\beta$ to infinity, $M_{UV}/M_{KK}\to +\infty$, therefore instanton particles become heavy and are integrated out.

In the limit we are considering we scale also the Wilson loop $\Uns$ as\footnote{This should correspond to the expansion in $\beta$ of the Wilson loop observable
\begin{equation*}
\ev{\Tr_2 P\text{-}\exp\int_{S^1_\beta} \phi(x,t) \dd t}_{\mathbb{C}^2\times S^1_\beta,NS}=\ev{\Tr_2\left(1+\beta\phi(x)+\frac{1}{2}\beta^2\phi^2(x)+O(\beta^3)\right)}_{\mathbb{C}^2,NS}\ , 
\end{equation*}
with
 \begin{equation*}
 \int_{S^1_\beta} \phi(x,t) \dd t =\beta\phi(x)\ ,  \quad \Tr_2 \phi(x)=0\ ,
 \end{equation*}
 and $\phi(x)$ the 4d complex scalar.}
\begin{equation}
\Uns=2+\left(\uns-\frac{\epsilon^2}{4}\right) \beta^2+O(\beta^3)=2+\left(\frac{\uns}{\epsilon^2}-\frac{1}{4}\right)(q-1)^2+O((q-1)^3) \ .
\end{equation}
The finite continuous 4d surface observable parameter $s$ is obtained by sending $d\to\infty$ as
\begin{equation}
d=\frac{\epsilon s}{q-1} \ 
\end{equation}
and we define the NS 4d blowup factor\footnote{The 4d limit is independent on the value of the CS level $k$. We also observe, from \eqref{sec2,1}, that the 4d and 5d blowup factors differ by a ``gauge prefactor'' $q^{-d/4}$.} as
\begin{equation}
\mathcal{B}_{NS}^{(j)}(s)=\lim_{\beta\to0}q^{-d/4}\mathcal{B}_{NS}^{(j,k)}\left(\frac{\epsilon s}{q-1},\beta\right)=\ev{e^{sO^{(4d)}(\Omega,{\bf F}^2)}}_{\hat{\mathbb{C}}^2,NS} \ , 
\end{equation}
where $O^{(4d)}(\Omega,{\bf F}^2)$ is the 4d surface observable \eqref{sec1,0.75} inserted in the exceptional divisor $E$. We now consider the expansion
\begin{align}
&B_{d+m}^{(j,k)}B_{d-m}^{(j,k)}={\mathcal{B}}^{(j,k)}_{NS}\left(d+m,\beta\right){\mathcal{B}}^{(j,k)}_{NS}\left(d-m,\beta\right)=e^{m D}({\mathcal{B}}^{(j,k)}_{NS},{\mathcal{B}}^{(j,k)}_{NS})= \nonumber\\
&=\sum_{n=0}^{\infty} \frac{m^n}{n!}\HirD{{\mathcal{B}}^{(j,k)}_{NS}}{d}{n}=\sum_{n=0}^{\infty} \frac{m^n}{n!}\left(\frac{q - 1}{\epsilon}\right)^{n}\HirD{{\mathcal{B}}^{(j,k)}_{NS}}{s}{n}\ .
\end{align}
In the limit $q\to1$ both the Hirota equations \eqref{sec3,3} and \eqref{sec3,7} become
\begin{equation}
D_s^{(2)}(\mathcal{B}_{NS}^{(j)},\mathcal{B}_{NS}^{(j)})=-2e^{\frac{\epsilon s}{2}} \Lambda^2 \, \mathcal{B}_{NS}^{(j-1)}\mathcal{B}_{NS}^{(j+1)}\ ,
\end{equation}
which is the differential Hirota Painlevé III$_3$ equation in Toda form, where $\Lambda^4$ parametrizes the position of the zero in the odd sector $j=1$. The surface observable parameter $s$ shifts the 4d coupling $\Lambda^4$ and the Hurwitz expansion is obtained expanding around $s=0$ \cite{Bonelli:2024wha}
\begin{equation}
\Lambda^4_{\epsilon s}=\Lambda^4 e^{\epsilon s}\ , \quad  \mathcal{B}_{NS}^{(j)}(s)=1-j+\sum_{n=0}^{+\infty} c_n^{PIII_3,(j)}\frac{s^{n+1}}{(n+1)!}\ ,
\end{equation}
where the coefficients\footnote{For the first values of the coefficients see appendix D of \cite{Bonelli:2024wha}.} are integer polynomials in $\uns$, $\Lambda^4$ and $\epsilon$.

Finally, the Hurwitz expansion around $s=0$ is easily recovered observing that the $q\to1$ limit of the discrete derivative $D_q$ at $d=0$ gives the continuous derivative $\partial_s$ evaluated at $s=0$ 
\begin{align}
&c_n^{PIII_3,(j)}=\epsilon^n\lim_{q\to 1} D_q^{n}(q^{-d/4}B^{(j,k)}_{d}(\Uns,z,q))\eval_{d=0}=\partial_s^n\mathcal{B}_{NS}^{(j)}(\uns,\Lambda,s)\eval_{s=0} \ , \\
&D_q B^{(j,k)}_{d}(z)\equiv\frac{B^{(j,k)}_{d+1}(z)-B^{(j,k)}_{d}(z)}{q-1}\ .
\end{align}

 \subsection{Strongly coupled 4d limit at negative coupling}\label{5dsection4.2}
 We consider now the 4d limit to H$_{0}$ AD theory. This is achieved considering a double scaling limit of the $k=1$ theory in which we keep finite the UV coupling $z \leftrightarrow 1/g^2_
4(1/\beta)$, i.e. the mass ratio $M_{UV}/M_{KK}=\beta/g_5^2=-\log z$ is kept fixed in the limit $\beta \to 0$.

To get the limit to the AD theory we have to tune the coupling $z$ and the Wilson loop $\Uns$ to some special finite values. The deformations of the theory at this point have exactly the correct fractional scaling dimensions of the AD SCFT. In the following we will illustrate in detail how this scaling arises from the $q$-Painlevé equation. Furthermore, we find that the first Chern class $j$ cannot be kept fixed in the flow of the $5d$ surface observable to the IR. At the level of the equations this means that we actually have to consider an alternating dynamics where we jump from $\Tau^{(0,1)}$ to $\Tau^{(1,1)}$ depending on the value of $d$. 

Precisely, we now set $k=1$ and we define the ``twisted'' NS blowup factor
\begin{equation}
 \tilde{\mathcal{B}}_{NS}(a,\Lm,\epsilon,d,\beta)=\tilde B_{d}(\Uns,z,q)\equiv
 \begin{cases}
  B_{d/2}^{(1,1)}(\Uns,z,q) \quad &\text{ if $d$ even}\ ,\\
  B_{d/2+1/2}^{(0,1)}(\Uns,z,q) \quad &\text{ if $d$ odd}\ ,
 \end{cases}
\end{equation}
and we can rewrite the equation \eqref{sec3,7} in terms\footnote{The equation \eqref{sec3,9} corresponds at even times to \eqref{sec3,7} for $j=1$ and at odd times to \eqref{sec3,7} for $j=0$.}  of $\tilde B_{d}$
\begin{equation}\label{sec3,9}
\tilde B_{d+2}\tilde B_{d-2} = \tilde B_d\tilde B_d  - z^{1/2}q^{d/4} \tilde B_{d+1}\tilde B_{d-1}\ .
\end{equation}
We want now  to recover the AD theory which corresponds to the PI equation
\begin{equation}
\HirD{\mathcal{B}_{NS}}{s}{4}+2\epsilon s \mathcal{B}_{NS}^2 -g_2\mathcal{B}_{NS}^2= 0 \ .
\end{equation} 
To do this we scale the parameters of the theory in the limit $q\to1$ as
 \begin{equation}\label{sec3,13}
d = -\delta^{-1}s\ ,\quad z=z_0\left(1 + \frac{g_2}{4} \delta^p\right)\ , \quad \delta= \left(\frac{q-1}{\epsilon}\right)^{\Delta}\ ,
 \end{equation}
The scaling dimensions $\Delta$ and $\Delta'=p\Delta$ correspond precisely to the ones of the surface observable parameter $s$ and of the gauge coupling $g_2$ of the 4d theory we reach in the limit. We define now
\begin{equation}\label{sec3,10}
\tilde{\mathcal{B}'}_{NS}\left(d,\beta\right) =   \kappa^{-d^2} \tilde B_d\ ,
\end{equation}
and the 4d blowup factor
 \begin{equation}
\tilde{\mathcal{B}}^{(4d)}_{NS}(s)=\lim_{\beta\to0}\tilde{\mathcal{B}}'_{NS}\left(d,\beta\right) \ .  
\end{equation}
As we will show below, the blowup factor $\tilde{\mathcal{B}}_{NS}^{(4d)}(s)$ is precisely the one of AD theory H$_{0}$. The gaussian prefactor contribution $\kappa$ in \eqref{sec3,10} is the contribution to the contact term $T(\Uns)$ \eqref{sec2,6.5} of the dof that decouple in the AD limit. 

To take the limit we rewrite
\begin{align}\label{sec3,11}
&\tilde B_{d+m}\tilde B_{d-m}=\kappa^{2d^2+2m^2}\tilde{\mathcal{B}}'_{NS}\left(d+m,\beta\right)\tilde{\mathcal{B}}'_{NS}\left(d-m,\beta\right)\ , \nonumber\\
&\tilde{\mathcal{B}'}_{NS}\left(d+m,\beta\right)\tilde{\mathcal{B}'}_{NS}\left(d-m,\beta\right)=e^{m D}(\tilde{\mathcal{B}}'_{NS},\tilde{\mathcal{B}}'_{NS})= \nonumber\\
&=\sum_{n=0}^{\infty} \frac{m^n}{n!}\HirD{\tilde{\mathcal{B}}'_{NS}}{d}{n}=\sum_{n=0}^{\infty} \frac{m^n}{n!}\delta^{n}\HirD{\tilde{\mathcal{B}}'_{NS}}{s}{n}\ ,
\end{align}
where $\HirD{\tilde{\mathcal{B}}_{NS}}{d}{n}$ is the $n$-th Hirota derivative and where in the last step we changed variable $d\to s$.

Rewriting the equation \eqref{sec3,9} in terms of $\tilde{\mathcal{B}}'_{NS}$ and dividing by $\kappa^{2d^2+8}$ we get
\begin{equation}\label{sec3,12}
\tilde{\mathcal{B}}'_{NS}\left(d+2,\beta\right)\tilde{\mathcal{B}}'_{NS}\left(d-2,\beta\right) = \kappa^{-8}\tilde{\mathcal{B}}'_{NS}\left(d,\beta\right)^2- z^{1/2}q^{d/4}\kappa^{-6} \tilde{\mathcal{B}}'_{NS}\left(d+1,\beta\right)\tilde{\mathcal{B}}'_{NS}\left(d-1,\beta\right)\ .
\end{equation}
 and expanding the equation \eqref{sec3,12} for $q\to1$ and using \eqref{sec3,13}, \eqref{sec3,11} we obtain\footnote{Notice that $\HirD{\tilde{\mathcal{B}}'_{NS}}{s}{n}$ is vanishing for $n$ odd.}
\begin{align}\label{sec3,14}
&\left(1 + z_0^{1/2} \kappa^{-6} - \kappa^{-8} \right)\tilde{\mathcal{B}}'_{NS}(s)^2+ \nonumber\\
&\left(2+\frac{1}{2}z_0^{1/2} \kappa^{-6} \right)\HirD{\tilde{\mathcal{B}}'_{NS}}{s}{2}\delta^2+ \nonumber \\
&\frac{1}{24} \left(16 + z_0^{1/2} \kappa^{-6} \right)\HirD{\tilde{\mathcal{B}}'_{NS}}{s}{4}\delta^4+\nonumber\\ 
&\frac{1}{8}g_2 z_0^{1/2}\kappa^{-6} \tilde{\mathcal{B}}'_{NS}(s)^2 \delta^p -\frac{1}{4}\epsilon s z_0^{1/2}\kappa^{-6} \tilde{\mathcal{B}}'_{NS}(s)^2\delta^{\frac{1}{\Delta}-1}+\dots \ ,
\end{align}
where $\dots$ are subleading terms which do not contribute in the limit $q\to1$. The dynamics of the 4d theory is determined by the first non-zero term in the $\delta$ expansion and depends on the values of the parameters $\Delta,p,z_0,\kappa$.

\begin{figure}
    \centering
   \begin{tikzpicture}
\fill[black!5] (0,0) rectangle (6,6);

 \fill[pattern=north east lines, pattern color=black]
    (0,2) rectangle (1.65,6);

\definecolor{skyblue}{RGB}{156, 243, 255}
\definecolor{orange2}{RGB}{255, 138, 0}
\fill[skyblue] (0,4) rectangle (1,6);
\draw[->] (0,0) -- (6,0) node[right] {$\Delta$};
\draw[->] (0,0) -- (0,6) node[above] {$p$};
\draw[thick,dashed] (1.65,2) -- (1.65,6);
\draw[thick,dashed] (5,2) -- (5,6);
\draw[thick,dashed] (0,2) -- (1.65,2);
\draw[line width=1.5pt,orange2] (1,4) -- (1,6);
\draw[line width=1.5pt,blue] (0,4) -- (1,4);
 \fill[red] (1,4) circle (2.2pt) node[above left] {H$_0$};
 \foreach \x in {0,2,4}
    \draw (0.1,\x) -- (-0.1,\x) node[left] {\x};
  
    \draw (1,0.1) -- (1,-0.1) node[below] {$\frac{1}{5}$};
    \draw (1.65,0.1) -- (1.65,-0.1) node[below] {$\frac{1}{3}$};
    \draw (5,0.1) -- (5,-0.1) node[below] {1};

\end{tikzpicture}
    \caption{ \label{fig4:phasediagram}Phase diagram for the 4d scaling limits. The grey region corresponds to the trivial theory $\Tau=0$ and the dashed regions correspond to gapped theories ($\Tau\sim e^{f(s)}$) that we obtain only for generic values of $z_0$. The red point corresponds to the H$_0$ AD SCFT, the orange line to H$_0$ with no coupling deformation and the blue line and the light blue region are the SW limits of the previous two.}
\end{figure}
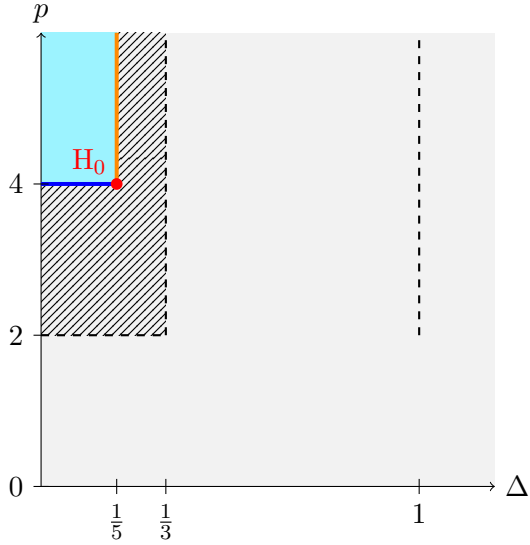

Looking at the leading contribution in $\delta$ in the equation \eqref{sec3,14} as we move in the parameter space we can construct a full phase diagram, which is reported in figure \ref{fig4:phasediagram}. For a generic value of the coupling\footnote{We can always fix $\kappa$ in terms of $z_0$ to cancel the zero order contribution in \eqref{sec3,14} which otherwise always trivializes the dynamics.} $z_0$ the theory is trivial ($\Tau=0$), except for $p\ge 2$ together with $\Delta=1$ or $\Delta\le 1/3$ where a gapped phase arises, shown as dashed line and dashed region in figure \ref{fig4:phasediagram}.
In this phase the only surviving contribution is the one of the topological defect coming from some contact term ($\Tau\sim e^{f(s)}$ with $f(s)$ some elementary function).

For a special value of $z_0$ instead we obtain the AD theory. The corresponding PI equation can appear only at order $\delta^4$ therefore the contributions at lower orders must cancel and the extra terms at order $\delta^p,\delta^{\frac{1}{\Delta}-1}$ must be of order $\delta^4$ . This happens if we have
\begin{equation}
\kappa^8= -\frac{1}{3} \ , \quad z_0^{1/2}\kappa^{-6}=-4\ \ , \quad \Delta=\frac{1}{5}\ , \quad p=4 \Rightarrow\Delta'=\frac{4}{5}\ ,
\end{equation}
which reproduce the correct scaling dimensions of the H$_0$ theory operators. In particular, we have
\begin{equation}
|\kappa|=\left(\frac{1}{3}\right)^{1/8}, \quad |z_0|=16\left(\frac{1}{3}\right)^{3/2}>1\ ,  
\end{equation}
therefore the AD point appear in the negative coupling phase of the $k=1$ theory corresponding to a deformation of the local $\mathbb{P}^2$ geometry.

Using this, in the limit $q\to1$ the equation \eqref{sec3,14} becomes exactly the quartic differential Hirota bilinear equation for the AD blowup factor
\begin{equation}
\HirD{\tilde{\mathcal{B}}_{NS}}{s}{4}+2\epsilon s \tilde{\mathcal{B}}_{NS}^2 -g_2\tilde{\mathcal{B}}_{NS}^2= 0 \ ,
\end{equation}   
which corresponds to the equation for the PI $\Tau$-function. 
If the deformation $g_2$ scales too fast ($p>4$) it is irrelevant in the limit and we obtain the AD theory where only the Coulomb branch operator deformation is present i.e. effectively $g_2=0$. If instead the surface observable, scales too slow $\Delta<1/5$ then the radius $\beta$ scales faster than the $\Omega$-background parameter $\epsilon$ and it washes it out leading just to the SW limit of the theory, the $\Tau$-function is given then by \eqref{sec2,2.5}.

We observe that this behaviour is possible precisely because of the presence of the CS term which introduces the extra shift in the term $\tilde{\mathcal{B}}'_{NS}\left(d+1,\beta\right)\tilde{\mathcal{B}}'_{NS}\left(d-1,\beta\right)$. This breaks the symmetry $z\to 1/z$ and gives an extra $\HirD{\tilde{\mathcal{B}}'_{NS}}{s}{2}$ contribution that allows for the cancellation of the $\delta^2$ term. Conversely, for $k=0$ no cancellation arises and we don't find any point with non-trivial dynamics as expected from the analysis of the geometries we discussed before.

    We can also take the limit directly at the level of the $\Tau$-function scaling the Wilson loop parameter as follows\footnote{This particular scaling is obtained matching the solution with the Hurwitz expansion of the 4d AD theory and should be understood as the RG flow of the Wilson loop observable to the IR AD point.}
\begin{equation}\label{sec3,8}
\Uns=-\frac{2i}{3 \sqrt{3}} \left( 1 - \frac{g_2}{2}\delta^4 - 4 \epsilon \delta^5 - \frac{3}{2} {\bf g_3} \delta^6 \right)\ ,
\end{equation}
where $g_2$ is the AD coupling and ${\bf g_3}=-2\uns$ is the NS modulus corresponding to the Coulomb branch operator, reproduces the correct scaling dimension $6/5$ of H$_0$ theory and also contain some $\epsilon$ corrections which are due to the NS quantum SW geometry\footnote{In the SW limit $\epsilon\to 0$, the same scalings of $z,U$ were found in \cite{Closset:2021lhd} studying the SW $U$-plane of 5d susy gauge theories. This is consistent with the autonomous limit of the $q$-Painlevé equation which corresponds to the SW theory.}.

The shifted coupling of AD theory is additive $g_2(s)=g_2+\epsilon^{1/5} s$ and the surface observable parameter has fractional dimension $[s]=-1/5$. The Hurwitz expansion\footnote{See appendix D of \cite{Bonelli:2024wha} for the first values of the coefficients.} around $s=0$ 
\begin{equation}
 \tilde{\mathcal{B}}_{NS}(s)=\sum_{n=0}^{+\infty} c_n^{PI}\frac{s^{n+1}}{(n+1)!}\ ,  
\end{equation}
can be recovered taking the continuous limit of the ``fractional'' discrete derivatives $\tilde D_q^n$ of the 5d blowup factor
\begin{align}
 &c_n^{PI}=\epsilon^{n/5}\lim_{q\to 1} \tilde D_q^{n}\left[\left(-\frac{1}{3}\right)^{-d^2/8}\tilde B_{d}(\Uns,z,q)\right]\eval_{d=0}=\partial_s^n\tilde{\mathcal{B}}_{NS}(\uns,g_2,s)\eval_{s=0} \ , \\
& \tilde D_q \tilde B_{d}(z)\equiv\frac{\tilde B_{d+1}(z)-\tilde B_{d}(z)}{(q-1)^{1/5}}\ . 
\end{align}
In this way we recover the 4d $\Tau$-function, therefore the self-dual partition function, of the AD theory.

\newpage
\appendix
\phantom{some space}
\vfill
\section{The first coefficients of the Hurwitz expansions}\label{appA}
In this appendix we give the coefficients of the Hurwitz expansions studied in section \ref{5dsection3}. We computed them up to the order $d=15$ but for simplicity we report only the first coefficients ($0\leq d\leq6$). We recall that the NS blowup factor \eqref{sec2,6} has the following structure
\begin{align}
&\mathcal{B}_{NS}^{(j,k)}(a,\Lm,\epsilon,d,\beta)\equiv B^{(j,k)}_{d}(\Uns,z,q)=(q^d z)^{\frac{j}{4}}P^{(j,k)}_{d}(\Uns,z,q)=(q^d z)^{\frac{j}{4}}\sum_{n=0}^{n_{\max}^{(j,k)}(d)}P_{d,n}^{(j,k)}(z,q) \Uns^n \ .
\end{align}
where $j$ is the first Chern class of the blowup, $k$ is the CS level and $d$ is the number of insertions of the codimension 2 observable $I(E)$, and we have
\begin{align}
&\Uns(a,z,q^{-1})= \Uns(a,z,q)\ , &\text{ for }k=0\ , \nonumber\\
&\tilde \Uns(a,z,q)\equiv\Uns(a,z,q^{-1}) = \Uns(a,z,q) + (q^{-1/2} - q^{1/2}) z\ , &\text{ for }k=1\ , \nonumber \\
&B_{-d}^{(j,0)}(\Uns,z,q)=(-1)^jB_d^{(j,0)}(\Uns,z,q)\ , \quad B_{-d}^{(j,1)}(\Uns,z,q)=(-1)^jB_d^{(j,1)}(\tilde\Uns,z,q)\ .
\end{align}
In the following we report the $q$-polynomials $P^{(j,k)}_{d}(\Uns,z,q)$ in the NS Wilson loop variable $\Uns=\ev{W_1}_{NS}$ and the coupling $z$.

\subsection{$q$-PIII$_3$}\label{appA.1}
In this subsection we report the $P_{d}^{(j,0)}$ for the recursion \eqref{sec3,4} of $q$-PIII$_3$ ($k=0$) in the two sectors $j=0,1$. We find $n_{\max}^{(j,0)}(d) = \left\lfloor \frac{d^2}{4} \right\rfloor-\chi_4(d+2-2j)$ %\footnote{We have also $P_{d,n}^{(j,0)}=0$ if $n=1-j(d+1)\mod 2$.}.
\begin{align*}
P_{0}^{(0,0)}&=1\ ,\\[-3pt]
\\[-3pt]
P_{1}^{(0,0)}&=1\ ,\\[-3pt]
\\[-3pt]
P_{2}^{(0,0)}&=1-q z\ ,\\[-3pt]
\\[-3pt]
P_{3}^{(0,0)}&=(1-2 q z+q^2 z^2)-\\[0.5pt]
&\quad \Uns^2 q^2 z \ ,\\[-3pt]
\\[-3pt]
P_{4}^{(0,0)}&=1+(-3 q-q^3) z+(3 q^2+3 q^4) z^2+(-q^3-3 q^5) z^3+q^6 z^4+\\[0.5pt]
&\quad\Uns^2\big[(-2 q^2+2 q^3) z+(2 q^3-2 q^4) z^2\big]-\\[0.5pt]
&\quad\Uns^4q^3 z \ ,\\[-3pt]
\\[-3pt]
P_{5}^{(0,0)}&=1+(-4 q-2 q^3) z+(6 q^2+8 q^4+q^6) z^2+(-4 q^3-12 q^5-4 q^7) z^3+(q^4+8 q^6+6 q^8) z^4+\\
&\qquad(-2 q^7-4 q^9) z^5+q^{10} z^6+\\[0.5pt]
&\quad\Uns^2\big[ (-3 q^2+4 q^3-4 q^4) z+(6 q^3-8 q^4+10 q^5+4 q^6) z^2+(-3 q^4+4 q^5-8 q^6-8 q^7-3 q^8) z^3+\\
&\qquad(2 q^7+4 q^8+6 q^9) z^4-3 q^{10} z^5\big]+\\[0.5pt]
&\quad\Uns^4\big[(-2 q^3+4 q^4) z+(3 q^4-4 q^5-2 q^6) z^2+q^{10} z^4\big]-\\
&\quad\Uns^6 q^4 z \ ,\\[-3pt]
\\[-3pt]
P_{6}^{(0,0)}&=1+(-5 q-3 q^3-q^5) z+(10 q^2+15 q^4+8 q^6+3 q^8) z^2+(-10 q^3-30 q^5-25 q^7-16 q^9-3 q^{11}) z^3+\\
&\qquad(5 q^4+30 q^6+40 q^8+35 q^{10}+15 q^{12}+q^{14}) z^4+(-q^5-15 q^7-35 q^9-40 q^{11}-30 q^{13}-5 q^{15}) z^5+\\
&\qquad(3 q^8+16 q^{10}+25 q^{12}+30 q^{14}+10 q^{16}) z^6+(-3 q^{11}-8 q^{13}-15 q^{15}-10 q^{17}) z^7+\\
&\qquad(q^{14}+3 q^{16}+5 q^{18}) z^8-q^{19} z^9+\\[0.5pt]
&\quad\Uns^2\big[(-4 q^2+6 q^3-8 q^4+6 q^5) z+(12 q^3-18 q^4+30 q^5-14 q^6-2 q^7-8 q^8) z^2+\\
&\qquad(-12 q^4+18 q^5-42 q^6+6 q^7-2 q^8+22 q^9+4 q^{10}+6 q^{11}) z^3+\\[0.5pt]
&\qquad(4 q^5-6 q^6+26 q^7+6 q^8+18 q^9-18 q^{10}-6 q^{11}-26 q^{12}+6 q^{13}-4 q^{14}) z^4+\\
&\qquad(-6 q^8-4 q^9-22 q^{10}+2 q^{11}-6 q^{12}+42 q^{13}-18 q^{14}+12 q^{15}) z^5+\\
&\qquad(8 q^{11}+2 q^{12}+14 q^{13}-30 q^{14}+18 q^{15}-12 q^{16}) z^6+(-6 q^{14}+8 q^{15}-6 q^{16}+4 q^{17}) z^7\big]+\\[0.5pt]
&\quad\Uns^4\big[ (-3 q^3+8 q^4-11 q^5) z+(9 q^4-20 q^5+21 q^6+12 q^7+8 q^8) z^2+\\
&\qquad(-6 q^5+12 q^6-13 q^7-20 q^8-22 q^9-8 q^{10}-3 q^{11}) z^3+\\
&\qquad(3 q^8+8 q^9+22 q^{10}+20 q^{11}+13 q^{12}-12 q^{13}+6 q^{14} ) z^4+\\
&\qquad(-8 q^{11}-12 q^{12}-21 q^{13}+20 q^{14}-9 q^{15}) z^5+(11 q^{14}-8 q^{15}+3 q^{16}) z^6\big]+\\[0.5pt]
&\quad\Uns^6\big[(-2 q^4+6 q^5) z+(4 q^5-6 q^6-4 q^7-2 q^8) z^2+(2 q^{11}+4 q^{12}+6 q^{13}-4 q^{14}) z^4+\\
&\qquad(-6 q^{14}+2 q^{15}) z^5\big]+\\[0.5pt]
&\quad\Uns^8(-q^5 z+q^{14} z^4)\ ,
\end{align*}

\begin{align*}
P_{0}^{(1,0)}&=0\ ,\\
\\[-3pt]
P_{1}^{(1,0)}&=-1\ ,\\
\\[-3pt]
P_{2}^{(1,0)}&=-\Uns\ ,\\
\\[-3pt]
P_{3}^{(1,0)}&=1-2 q z+q^2 z^2-\\
&\quad\Uns^2\ ,\\
\\[-3pt]
P_{4}^{(1,0)}&=\Uns\big[2+(-4 q-2 q^2) z+(2 q^2+4 q^3) z^2-2 q^4 z^3\big]+\\
&\quad\Uns^3\big[-1+q^4 z^2\big]\ ,\\
\\[-3pt]
P_{5}^{(1,0)}&=-1+(4 q+2 q^3) z+(-6 q^2-8 q^4-q^6) z^2+(4 q^3+12 q^5+4 q^7) z^3+(-q^4-8 q^6-6 q^8) z^4+\\
&\qquad(2 q^7+4 q^9) z^5-q^{10} z^6+\\
&\quad\Uns^2\big[3+(-6 q-4 q^2-2 q^3) z+(3 q^2+8 q^3+8 q^4-4 q^5+3 q^6) z^2+(-4 q^4-10 q^5+8 q^6-6 q^7) z^3+\\
&\qquad(4 q^6-4 q^7+3 q^8) z^4\big]+\\
&\quad\Uns^4\big[-1+(2 q^4+4 q^5-3 q^6) z^2+(-4 q^6+2 q^7) z^3\big]+\\
&\quad\Uns^6 q^6 z^2\ ,\\
\\[-3pt]
P_{6}^{(1,0)}&=\Uns\big[-3+(12 q+4 q^2+4 q^3+4 q^4) z+(-18 q^2-16 q^3-16 q^4-24 q^5-2 q^6-8 q^7) z^2+\\
&\qquad(12 q^3+24 q^4+24 q^5+56 q^6+8 q^7+36 q^8+4 q^9+4 q^{10}) z^3+\\
&\qquad(-3 q^4-16 q^5-16 q^6-64 q^7-12 q^8-64 q^9-16 q^{10}-16 q^{11}-3 q^{12}) z^4+\\
&\qquad(4 q^6+4 q^7+36 q^8+8 q^9+56 q^{10}+24 q^{11}+24 q^{12}+12 q^{13}) z^5+\\
&\qquad(-8 q^9-2 q^{10}-24 q^{11}-16 q^{12}-16 q^{13}-18 q^{14}) z^6+(4 q^{12}+4 q^{13}+4 q^{14}+12 q^{15}) z^7-3 q^{16} z^8\big]+\\
&\quad\Uns^3\big[4+(-8 q-6 q^2-4 q^3-2 q^4) z+(4 q^2+12 q^3+14 q^4+4 q^5-10 q^6+20 q^7-8 q^8) z^2+\\
&\qquad(-6 q^4-16 q^5-10 q^6+40 q^7-50 q^8+16 q^9+6 q^{10}) z^3+\\
&\qquad(6 q^6+16 q^7-50 q^8+40 q^9-10 q^{10}-16 q^{11}-6 q^{12}) z^4+\\
&\qquad(-8 q^8+20 q^9-10 q^{10}+4 q^{11}+14 q^{12}+12 q^{13}+4 q^{14}) z^5+\\
&\qquad(-2 q^{12}-4 q^{13}-6 q^{14}-8 q^{15}) z^6+4 q^{16} z^7\big]+\\
&\quad\Uns^5\big[-1+(3 q^4+8 q^5+2 q^6-16 q^7+12 q^8) z^2+(-8 q^6-16 q^7+32 q^8-16 q^9-8 q^{10}) z^3+\\
&\qquad(12 q^8-16 q^9+2 q^{10}+8 q^{11}+3 q^{12}) z^4-q^{16} z^6\big]+\\
&\quad\Uns^7\big[(2 q^6+4 q^7-6 q^8) z^2+(-6 q^8+4 q^9+2 q^{10}) z^3\big]+\\
&\quad\Uns^9 q^8 z^2\ ,\\
\\[-3pt]
\end{align*} 

\subsection{$q$-PI}\label{appA.2}
In this subsection we report the $P_{d}^{(j,1)}$ for the recursion \eqref{sec3,7.25} of $q$-PI ($k=1$) in the two sectors $j=0,1$. We denote $\widetilde z =z q^{3/2}$ and we find  $n_{\max}^{(j,1)}(d) =$ {\scriptsize $\left\lfloor \frac{d(d+j-1)}{4} \right\rfloor -j\chi_4(d) $}
\begin{align*}
P_{0}^{(0,1)}&=1\ ,\\[-3pt] 
\\[-3pt]
P_{1}^{(0,1)}&=1\ ,\\[-3pt]
\\[-3pt]
P_{2}^{(0,1)}&=1\ ,\\
\\[-3pt]
P_{3}^{(0,1)}&=1-\\
&\quad\Uns{\widetilde z}\ ,\\[-3pt]
\\[-3pt]
P_{4}^{(0,1)}&=1+\\
&\quad\Uns(-2+q) {\widetilde z}+\\
&\quad\Uns^2(1-q) {\widetilde z}^2-\\
&\quad\Uns^3 q {\widetilde z}\ ,\\[-3pt]
\\[-3pt]
P_{5}^{(0,1)}&=1+(q^2-q^3) {\widetilde z}^2+\\
&\quad\Uns\big[(-3+2 q-2 q^2) {\widetilde z}+(-2 q^2+2 q^3) {\widetilde z}^3\big]+\\
&\quad\Uns^2\big[(3-4 q+2 q^2+2 q^3) {\widetilde z}^2+(q^2-q^3) {\widetilde z}^4\big]+\\
&\quad\Uns^3\big[(-2 q+3 q^2) {\widetilde z}+(-1+2 q-2 q^3) {\widetilde z}^3\big]+\\
&\quad\Uns^4(2 q-2 q^2-q^3) {\widetilde z}^2-\\
&\quad\Uns^5 q^2 {\widetilde z}\ ,\\[-3pt]
\\[-3pt]
P_{6}^{(0,1)}&=1+(2 q^2-3 q^3+q^4) {\widetilde z}^2+(q^3-2 q^4+2 q^6-q^7) {\widetilde z}^4\ ,\\
&\quad\Uns\big[(-4+3 q-4 q^2+2 q^3) {\widetilde z}+(-6 q^2+6 q^3+4 q^5-4 q^6) {\widetilde z}^3+(-3 q^3+6 q^4-6 q^6+3 q^7) {\widetilde z}^5\big]+\\
&\quad\Uns^2\big[(6-9 q+9 q^2+5 q^3-4 q^4-4 q^5) {\widetilde z}^2+(6 q^2-3 q^3-12 q^5+7 q^6+2 q^7) {\widetilde z}^4+\\
&\qquad (3 q^3-6 q^4+6 q^6-3 q^7) {\widetilde z}^6\big]+\\
&\quad\Uns^3\big[(-3 q+6 q^2-7 q^3) {\widetilde z}+(-4+9 q-6 q^2-13 q^3+2 q^4+5 q^5+6 q^6) {\widetilde z}^3+\\
&\qquad(-2 q^2-4 q^4+12 q^5-2 q^6-4 q^7) {\widetilde z}^5+(-q^3+2 q^4-2 q^6+q^7) {\widetilde z}^7\big]+\\
\end{align*}
\begin{align*}
&\quad\Uns^4\big[(6 q-12 q^2+5 q^3+6 q^4+4 q^5) {\widetilde z}^2+(1-3 q+q^2+6 q^3+2 q^4+2 q^5-8 q^6-q^7) {\widetilde z}^4+\\
&\qquad (3 q^4-4 q^5-q^6+2 q^7) {\widetilde z}^6\big]+\\
&\quad\Uns^5\big[(-2 q^2+5 q^3) {\widetilde z}+(-3 q+6 q^2+q^3-4 q^4-4 q^5-2 q^6) {\widetilde z}^3+(-3 q^5+2 q^6+q^7) {\widetilde z}^5\big]+\\
&\quad\Uns^6\big[(3 q^2-3 q^3-2 q^4-q^5) {\widetilde z}^2+q^6 {\widetilde z}^4\big]-\\
&\quad\Uns^7q^3 {\widetilde z}\ ,\\[-3pt]
\\[-3pt]     
\end{align*}
\begin{align*}
P_{0}^{(1,1)}&=0\ ,\\[-3pt]
\\[-3pt]
P_{1}^{(1,1)}&=-1\ ,\\[-3pt]
\\[-3pt]
P_{2}^{(1,1)}&=-\Uns\ ,\\[-3pt]
\\[-3pt]
P_{3}^{(1,1)}&=1-\\
&\quad\Uns{\widetilde z}-\\
&\quad\Uns^2\ ,\\[-3pt]
\\[-3pt]
P_{4}^{(1,1)}&=(-1+q) {\widetilde z}+\\
&\quad\Uns\big[2+(2-2 q) {\widetilde z}^2\big]+\\
&\quad\Uns^2\big[(-2-q) {\widetilde z}+(-1+q) {\widetilde z}^3\big]+\\
&\quad\Uns^3\big[-1+q {\widetilde z}^2\big]\ ,\\[-3pt]
\\[-3pt]
P_{5}^{(1,1)}&=-1+(1-2 q+q^3) {\widetilde z}^2+\\
&\quad\Uns\big[(q+2 q^2) {\widetilde z}+(-3+6 q-4 q^3+q^4) {\widetilde z}^3\big]+\\
&\quad\Uns^2\big[3+(3+q-8 q^2+q^3) {\widetilde z}^2+(3-6 q+5 q^3-2 q^4) {\widetilde z}^4\big]+\\
&\quad\Uns^3\big[(-3-2 q-q^2) {\widetilde z}+(-2-5 q+10 q^2-2 q^4) {\widetilde z}^3+(-1+2 q-2 q^3+q^4) {\widetilde z}^5\big]+\\
&\quad\Uns^4\big[-1+(2 q+4 q^2-3 q^3) {\widetilde z}^2+(3 q-4 q^2-q^3+2 q^4) {\widetilde z}^4\big]+\\
&\quad\Uns^5(-3 q^2+2 q^3+q^4) {\widetilde z}^3+\\
&\quad\Uns^6 q^3 {\widetilde z}^2\ ,\\[-3pt]
\end{align*}
\begin{align*}
P_{6}^{(1,1)}&=(2-2 q+q^2-q^3) {\widetilde z}+(-1+3 q-q^2-2 q^3+2 q^4-3 q^5+2 q^6) {\widetilde z}^3+(q^5-2 q^6+2 q^8-q^9) {\widetilde z}^5+\\
&\quad\Uns\big[-3+(-4+4 q-10 q^2+10 q^3-4 q^4+4 q^5) {\widetilde z}^2+\\
&\qquad(4-12 q+4 q^2+10 q^3-8 q^4+4 q^5-4 q^6+6 q^7-4 q^8) {\widetilde z}^4+(-3 q^5+6 q^6-6 q^8+3 q^9) {\widetilde z}^6\big]+\\
&\quad\Uns^2\big[(3+3 q+3 q^2+3 q^3) {\widetilde z}+(-5 q+36 q^2-31 q^3+3 q^4+3 q^5-2 q^6-4 q^7) {\widetilde z}^3+\\
&\qquad(-6+18 q-6 q^2-18 q^3+12 q^4+3 q^5+6 q^6-18 q^7+6 q^8+3 q^9) {\widetilde z}^5+(3 q^5-6 q^6+6 q^8-3 q^9) {\widetilde z}^7\big]+\\
&\quad\Uns^3\big[4+(3-8 q^2-20 q^3+16 q^4-10 q^5) {\widetilde z}^2+\\
&\qquad(4+11 q-58 q^2+37 q^3+18 q^4-15 q^5-6 q^6-q^7+10 q^8) {\widetilde z}^4+\\
&\qquad(4-12 q+4 q^2+14 q^3-8 q^4-6 q^5-8 q^6+18 q^7-6 q^9) {\widetilde z}^6+(-q^5+2 q^6-2 q^8+q^9) {\widetilde z}^8\big]+\\
&\quad\Uns^4\big[(-4-3 q-2 q^2-q^3) {\widetilde z}+(-3-9 q-3 q^2+55 q^3-35 q^4-2 q^5+4 q^6+8 q^7) {\widetilde z}^3+\\
&\qquad(-2-13 q+43 q^2-16 q^3-29 q^4+8 q^5+12 q^6+14 q^7-14 q^8-3 q^9) {\widetilde z}^5+\\
&\qquad(-1+3 q-q^2-4 q^3+2 q^4+2 q^5+4 q^6-6 q^7-2 q^8+3 q^9) {\widetilde z}^7\big]+\\
&\quad\Uns^5\big[-1+(3 q+8 q^2+2 q^3-16 q^4+12 q^5) {\widetilde z}^2+\\
&\qquad(6 q+18 q^2-56 q^3+16 q^4+22 q^5+4 q^6-8 q^7-8 q^8) {\widetilde z}^4+\\
&\qquad(5 q-12 q^2+q^3+12 q^4-4 q^6-9 q^7+4 q^8+3 q^9) {\widetilde z}^6\big]+\\
&\quad\Uns^6\big[(-6 q^2-13 q^3+31 q^4-5 q^5-6 q^6-5 q^7) {\widetilde z}^3+(-10 q^2+18 q^3+3 q^4-11 q^5-6 q^6+6 q^8+q^9) {\widetilde z}^5\big]+\\
&\quad\Uns^7\big[(2 q^3+4 q^4-6 q^5) {\widetilde z}^2+(10 q^3-12 q^4-5 q^5+2 q^6+4 q^7+2 q^8) {\widetilde z}^4\big]+\\
&\quad\Uns^8(-5 q^4+3 q^5+2 q^6+q^7) {\widetilde z}^3+\\
&\quad\Uns^9 q^5 {\widetilde z}^2\ ,\\[-3pt]
\\[-3pt]  
\end{align*}

\bibliographystyle{JHEP}
\bibliography{refPainlev}

\providecommand{\href}[2]{#2}\begingroup\raggedright\begin{thebibliography}{10}

\bibitem{Argyres:1995jj}
P.C.~Argyres and M.R.~Douglas, \emph{{New phenomena in SU(3) supersymmetric
  gauge theory}},
  \href{https://doi.org/10.1016/0550-3213(95)00281-V}{\emph{Nucl. Phys. B}
  {\bfseries 448} (1995) 93}
  [\href{https://arxiv.org/abs/hep-th/9505062}{{\ttfamily hep-th/9505062}}].

\bibitem{Argyres:1995xn}
P.C.~Argyres, M.R.~Plesser, N.~Seiberg and E.~Witten, \emph{{New N=2
  superconformal field theories in four-dimensions}},
  \href{https://doi.org/10.1016/0550-3213(95)00671-0}{\emph{Nucl. Phys. B}
  {\bfseries 461} (1996) 71}
  [\href{https://arxiv.org/abs/hep-th/9511154}{{\ttfamily hep-th/9511154}}].

\bibitem{Maruyoshi:2016tqk}
K.~Maruyoshi and J.~Song, \emph{{Enhancement of Supersymmetry via
  Renormalization Group Flow and the Superconformal Index}},
  \href{https://doi.org/10.1103/PhysRevLett.118.151602}{\emph{Phys. Rev. Lett.}
  {\bfseries 118} (2017) 151602}
  [\href{https://arxiv.org/abs/1606.05632}{{\ttfamily 1606.05632}}].

\bibitem{Maruyoshi:2016aim}
K.~Maruyoshi and J.~Song, \emph{{$ \mathcal{N}=1 $ deformations and RG flows of
  $ \mathcal{N}=2 $ SCFTs}},
  \href{https://doi.org/10.1007/JHEP02(2017)075}{\emph{JHEP} {\bfseries 02}
  (2017) 075} [\href{https://arxiv.org/abs/1607.04281}{{\ttfamily
  1607.04281}}].

\bibitem{Agarwal:2016pjo}
P.~Agarwal, K.~Maruyoshi and J.~Song, \emph{{$ \mathcal{N} $ =1 Deformations
  and RG flows of $ \mathcal{N} $ =2 SCFTs, part II: non-principal
  deformations}}, \href{https://doi.org/10.1007/JHEP12(2016)103}{\emph{JHEP}
  {\bfseries 12} (2016) 103}
  [\href{https://arxiv.org/abs/1610.05311}{{\ttfamily 1610.05311}}].

\bibitem{Agarwal:2017roi}
P.~Agarwal, A.~Sciarappa and J.~Song, \emph{{$ \mathcal{N} $ =1 Lagrangians for
  generalized Argyres-Douglas theories}},
  \href{https://doi.org/10.1007/JHEP10(2017)211}{\emph{JHEP} {\bfseries 10}
  (2017) 211} [\href{https://arxiv.org/abs/1707.04751}{{\ttfamily
  1707.04751}}].

\bibitem{Bonelli:2020dcp}
G.~Bonelli, F.~Del~Monte and A.~Tanzini, \emph{{BPS Quivers of Five-Dimensional
  SCFTs, Topological Strings and q-Painlev\'e Equations}},
  \href{https://doi.org/10.1007/s00023-021-01034-3}{\emph{Annales Henri
  Poincare} {\bfseries 22} (2021) 2721}
  [\href{https://arxiv.org/abs/2007.11596}{{\ttfamily 2007.11596}}].

\bibitem{Closset:2021lhd}
C.~Closset and H.~Magureanu, \emph{{The $U$-plane of rank-one 4d
  $\mathcal{N}=2$ KK theories}},
  \href{https://doi.org/10.21468/SciPostPhys.12.2.065}{\emph{SciPost Phys.}
  {\bfseries 12} (2022) 065}
  [\href{https://arxiv.org/abs/2107.03509}{{\ttfamily 2107.03509}}].

\bibitem{Seiberg:1996bd}
N.~Seiberg, \emph{{Five-dimensional SUSY field theories, nontrivial fixed
  points and string dynamics}},
  \href{https://doi.org/10.1016/S0370-2693(96)01215-4}{\emph{Phys. Lett. B}
  {\bfseries 388} (1996) 753}
  [\href{https://arxiv.org/abs/hep-th/9608111}{{\ttfamily hep-th/9608111}}].

\bibitem{Minahan:1996fg}
J.A.~Minahan and D.~Nemeschansky, \emph{{An N=2 superconformal fixed point with
  E(6) global symmetry}},
  \href{https://doi.org/10.1016/S0550-3213(96)00552-4}{\emph{Nucl. Phys. B}
  {\bfseries 482} (1996) 142}
  [\href{https://arxiv.org/abs/hep-th/9608047}{{\ttfamily hep-th/9608047}}].

\bibitem{boalch2007quivers}
P.~Boalch, \emph{{Quivers and difference Painlev\'e equations}},
  \href{https://arxiv.org/abs/0706.2634}{{\ttfamily 0706.2634}}.

\bibitem{Sakai2001RationalSA}
H.~Sakai, \emph{{Rational Surfaces Associated with Affine Root Systems and
  Geometry of the Painlev\'e Equations}},
  \href{https://doi.org/10.1007/s002200100446}{\emph{Communications in
  Mathematical Physics} {\bfseries 220} (2001) 165}.

\bibitem{Gamayun:2013auu}
O.~Gamayun, N.~Iorgov and O.~Lisovyy, \emph{{How instanton combinatorics solves
  Painlev\'e VI, V and IIIs}},
  \href{https://doi.org/10.1088/1751-8113/46/33/335203}{\emph{J. Phys. A}
  {\bfseries 46} (2013) 335203}
  [\href{https://arxiv.org/abs/1302.1832}{{\ttfamily 1302.1832}}].

\bibitem{Bonelli:2016qwg}
G.~Bonelli, O.~Lisovyy, K.~Maruyoshi, A.~Sciarappa and A.~Tanzini, \emph{{On
  Painlev\'e/gauge theory correspondence}},
  \href{https://doi.org/10.1007/s11005-017-0983-6}{\emph{Lett. Math. Phys.}
  {\bfseries 107} (2017) pages 2359}
  [\href{https://arxiv.org/abs/1612.06235}{{\ttfamily 1612.06235}}].

\bibitem{Gavrylenko:2016zlf}
P.~Gavrylenko and O.~Lisovyy, \emph{{Fredholm Determinant and Nekrasov Sum
  Representations of Isomonodromic Tau Functions}},
  \href{https://doi.org/10.1007/s00220-018-3224-7}{\emph{Commun. Math. Phys.}
  {\bfseries 363} (2018) 1} [\href{https://arxiv.org/abs/1608.00958}{{\ttfamily
  1608.00958}}].

\bibitem{Bonelli:2019boe}
G.~Bonelli, F.~Del~Monte, P.~Gavrylenko and A.~Tanzini, \emph{{${\mathcal {N}}$
  = $2^*$ Gauge Theory, Free Fermions on the Torus and Painlev\'e VI}},
  \href{https://doi.org/10.1007/s00220-020-03743-y}{\emph{Commun. Math. Phys.}
  {\bfseries 377} (2020) 1381}
  [\href{https://arxiv.org/abs/1901.10497}{{\ttfamily 1901.10497}}].

\bibitem{Bonelli:2019yjd}
G.~Bonelli, F.~Del~Monte, P.~Gavrylenko and A.~Tanzini, \emph{{Circular quiver
  gauge theories, isomonodromic deformations and $W_N$ fermions on the torus}},
  \href{https://doi.org/10.1007/s11005-020-01343-4}{\emph{Lett. Math. Phys.}
  {\bfseries 111} (2021) 83}
  [\href{https://arxiv.org/abs/1909.07990}{{\ttfamily 1909.07990}}].

\bibitem{Bonelli:2021rrg}
G.~Bonelli, F.~Globlek and A.~Tanzini, \emph{{Counting Yang-Mills Instantons by
  Surface Operator Renormalization Group Flow}},
  \href{https://doi.org/10.1103/PhysRevLett.126.231602}{\emph{Phys. Rev. Lett.}
  {\bfseries 126} (2021) 231602}
  [\href{https://arxiv.org/abs/2102.01627}{{\ttfamily 2102.01627}}].

\bibitem{Bonelli:2022iob}
G.~Bonelli, F.~Globlek and A.~Tanzini, \emph{{Toda equations for surface
  defects in SYM and instanton counting for classical Lie groups}},
  \href{https://doi.org/10.1088/1751-8121/ac9e2a}{\emph{J. Phys. A} {\bfseries
  55} (2022) 454004} [\href{https://arxiv.org/abs/2206.13212}{{\ttfamily
  2206.13212}}].

\bibitem{DelMonte:2022nem}
F.~Del~Monte, H.~Desiraju and P.~Gavrylenko, \emph{{Monodromy dependence and
  symplectic geometry of isomonodromic tau functions on the torus}},
  \href{https://doi.org/10.1088/1751-8121/acdc6c}{\emph{J. Phys. A} {\bfseries
  56} (2023) 294002} [\href{https://arxiv.org/abs/2211.01139}{{\ttfamily
  2211.01139}}].

\bibitem{Gavrylenko_2019}
P.~Gavrylenko, N.~Iorgov and O.~Lisovyy, \emph{{Higher rank isomonodromic
  deformations and $W$-algebras}},
  \href{https://doi.org/10.1007/s11005-019-01207-6}{\emph{Lett. Math. Phys.}
  {\bfseries 110} (2019) 327}
  [\href{https://arxiv.org/abs/1801.09608}{{\ttfamily 1801.09608}}].

\bibitem{Bershtein:2016aef}
M.A.~Bershtein and A.I.~Shchechkin, \emph{{q-deformed Painlev\'e $\tau$
  function and q-deformed conformal blocks}},
  \href{https://doi.org/10.1088/1751-8121/aa5572}{\emph{J. Phys. A} {\bfseries
  50} (2017) 085202} [\href{https://arxiv.org/abs/1608.02566}{{\ttfamily
  1608.02566}}].

\bibitem{Bonelli:2017gdk}
G.~Bonelli, A.~Grassi and A.~Tanzini, \emph{{Quantum curves and $q$-deformed
  Painlev\'e equations}},
  \href{https://doi.org/10.1007/s11005-019-01174-y}{\emph{Lett. Math. Phys.}
  {\bfseries 109} (2019) 1961}
  [\href{https://arxiv.org/abs/1710.11603}{{\ttfamily 1710.11603}}].

\bibitem{Bershtein:2017swf}
M.~Bershtein, P.~Gavrylenko and A.~Marshakov, \emph{{Cluster integrable
  systems, $q$-Painlev\'e equations and their quantization}},
  \href{https://doi.org/10.1007/JHEP02(2018)077}{\emph{JHEP} {\bfseries 02}
  (2018) 077} [\href{https://arxiv.org/abs/1711.02063}{{\ttfamily
  1711.02063}}].

\bibitem{jimbo2017cftapproachqpainlevevi}
M.~Jimbo, H.~Nagoya and H.~Sakai, \emph{{CFT approach to the q-Painlev\'e VI
  equation}}, \href{https://doi.org/10.1093/integr/xyx009}{\emph{J. Integrab.
  Syst.} {\bfseries 2} (2017) 1}
  [\href{https://arxiv.org/abs/1706.01940}{{\ttfamily 1706.01940}}].

\bibitem{Bonelli:2024wha}
G.~Bonelli, P.~Gavrylenko, I.~Majtara and A.~Tanzini, \emph{{Surface
  observables in gauge theories, modular Painlev\'e tau functions and
  non-perturbative topological strings}},
  \href{https://arxiv.org/abs/2410.17868}{{\ttfamily 2410.17868}}.

\bibitem{Nakajima:2003pg}
H.~Nakajima and K.~Yoshioka, \emph{{Instanton counting on blowup. 1.}},
  \href{https://doi.org/10.1007/s00222-005-0444-1}{\emph{Invent. Math.}
  {\bfseries 162} (2005) 313}
  [\href{https://arxiv.org/abs/math/0306198}{{\ttfamily math/0306198}}].

\bibitem{Nekrasov:2009rc}
N.A.~Nekrasov and S.L.~Shatashvili, \emph{{Quantization of Integrable Systems
  and Four Dimensional Gauge Theories}},  in \emph{{16th International Congress
  on Mathematical Physics}}, pp.~265--289, 8, 2009,
  \href{https://doi.org/10.1142/9789814304634_0015}{DOI}
  [\href{https://arxiv.org/abs/0908.4052}{{\ttfamily 0908.4052}}].

\bibitem{Bonelli:2025bmt}
G.~Bonelli, P.~Gavrylenko, T.~Pedroni and A.~Tanzini, \emph{{Blowing-up the
  edge: connection formulae and stability chart of the Lam{\'e} equation}},
  \href{https://arxiv.org/abs/2507.04860}{{\ttfamily 2507.04860}}.

\bibitem{DelMonte:2021ytz}
F.~Del~Monte and P.~Longhi, \emph{{Quiver Symmetries and Wall-Crossing
  Invariance}}, \href{https://doi.org/10.1007/s00220-022-04515-6}{\emph{Commun.
  Math. Phys.} {\bfseries 398} (2023) 89}
  [\href{https://arxiv.org/abs/2107.14255}{{\ttfamily 2107.14255}}].

\bibitem{DelMonte:2022kxh}
F.~Del~Monte and P.~Longhi, \emph{{The threefold way to quantum periods: WKB,
  TBA equations and q-Painlev{\'e}}},
  \href{https://doi.org/10.21468/SciPostPhys.15.3.112}{\emph{SciPost Phys.}
  {\bfseries 15} (2023) 112}
  [\href{https://arxiv.org/abs/2207.07135}{{\ttfamily 2207.07135}}].

\bibitem{Martone:2021drm}
M.~Martone and G.~Zafrir, \emph{{On the compactification of 5d theories to
  4d}}, \href{https://doi.org/10.1007/JHEP08(2021)017}{\emph{JHEP} {\bfseries
  08} (2021) 017} [\href{https://arxiv.org/abs/2106.00686}{{\ttfamily
  2106.00686}}].

\bibitem{Wang:2023zcb}
X.~Wang, \emph{{Wilson loops, holomorphic anomaly equations and blowup
  equations}},  \href{https://arxiv.org/abs/2305.09171}{{\ttfamily
  2305.09171}}.

\bibitem{hone2013properties}
A.N.W.~Hone, O.~Ragnisco and F.~Zullo, \emph{{Properties of the series solution
  for Painlevé I}},
  \href{https://doi.org/10.1080/14029251.2013.862436}{\emph{Journal of
  Nonlinear Mathematical Physics} {\bfseries 20} (2013) 85}
  [\href{https://arxiv.org/abs/1210.6822}{{\ttfamily 1210.6822}}].

\bibitem{hone2017hirota}
A.N.W.~Hone and F.~Zullo, \emph{{A Hirota bilinear equation for Painlevé
  transcendents PIV, PII and PI}},
  \href{https://doi.org/10.1142/S2010326318400014}{\emph{Random Matrices:
  Theory and Applications} {\bfseries 07} (2018) 1840001}
  [\href{https://arxiv.org/abs/1706.02341}{{\ttfamily 1706.02341}}].

\bibitem{Hori:2003ic}
K.~Hori, S.~Katz, A.~Klemm, R.~Pandharipande, R.~Thomas, C.~Vafa et~al.,
  \emph{{Mirror symmetry}}, vol.~1 of \emph{Clay mathematics monographs}, AMS,
  Providence, USA (2003).

\bibitem{Aharony:1997bh}
O.~Aharony, A.~Hanany and B.~Kol, \emph{{Webs of (p,q) five-branes,
  five-dimensional field theories and grid diagrams}},
  \href{https://doi.org/10.1088/1126-6708/1998/01/002}{\emph{JHEP} {\bfseries
  01} (1998) 002} [\href{https://arxiv.org/abs/hep-th/9710116}{{\ttfamily
  hep-th/9710116}}].

\bibitem{Bergman:2013ala}
O.~Bergman, D.~Rodr{\'\i}guez-G{\'o}mez and G.~Zafrir, \emph{{Discrete $\theta$
  and the 5d superconformal index}},
  \href{https://doi.org/10.1007/JHEP01(2014)079}{\emph{JHEP} {\bfseries 01}
  (2014) 079} [\href{https://arxiv.org/abs/1310.2150}{{\ttfamily 1310.2150}}].

\bibitem{Bergman:2013aca}
O.~Bergman, D.~Rodr{\'\i}guez-G{\'o}mez and G.~Zafrir, \emph{{5-Brane Webs,
  Symmetry Enhancement, and Duality in 5d Supersymmetric Gauge Theory}},
  \href{https://doi.org/10.1007/JHEP03(2014)112}{\emph{JHEP} {\bfseries 03}
  (2014) 112} [\href{https://arxiv.org/abs/1311.4199}{{\ttfamily 1311.4199}}].

\bibitem{Bridgeland:2024saj}
T.~Bridgeland, F.~Del~Monte and L.~Giovenzana, \emph{{Invariant Stability
  Conditions on Certain Calabi-Yau Threefolds}},
  \href{https://arxiv.org/abs/2412.08531}{{\ttfamily 2412.08531}}.

\bibitem{DelMonte:2023vwv}
F.~Del~Monte, \emph{{BPS Spectra and Algebraic Solutions of Discrete Integrable
  Systems}}, \href{https://doi.org/10.1007/s00220-024-05016-4}{\emph{Commun.
  Math. Phys.} {\bfseries 405} (2024) 147}
  [\href{https://arxiv.org/abs/2306.01626}{{\ttfamily 2306.01626}}].

\bibitem{Grassi:2014uua}
A.~Grassi, Y.~Hatsuda and M.~Marino, \emph{{Quantization conditions and
  functional equations in ABJ(M) theories}},
  \href{https://doi.org/10.1088/1751-8113/49/11/115401}{\emph{J. Phys. A}
  {\bfseries 49} (2016) 115401}
  [\href{https://arxiv.org/abs/1410.7658}{{\ttfamily 1410.7658}}].

\bibitem{Grassi:2014zfa}
A.~Grassi, Y.~Hatsuda and M.~Marino, \emph{{Topological Strings from Quantum
  Mechanics}}, \href{https://doi.org/10.1007/s00023-016-0479-4}{\emph{Annales
  Henri Poincare} {\bfseries 17} (2016) 3177}
  [\href{https://arxiv.org/abs/1410.3382}{{\ttfamily 1410.3382}}].

\bibitem{Moore:1997pc}
G.W.~Moore and E.~Witten, \emph{{Integration over the u plane in Donaldson
  theory}}, \href{https://doi.org/10.4310/ATMP.1997.v1.n2.a7}{\emph{Adv. Theor.
  Math. Phys.} {\bfseries 1} (1997) 298}
  [\href{https://arxiv.org/abs/hep-th/9709193}{{\ttfamily hep-th/9709193}}].

\bibitem{Manschot:2021qqe}
J.~Manschot and G.W.~Moore, \emph{{Topological correlators of $SU(2) N=2$ SYM
  on four-manifolds}},
  \href{https://doi.org/10.4310/atmp.240914021307}{\emph{Adv. Theor. Math.
  Phys.} {\bfseries 28} (2024) 407}
  [\href{https://arxiv.org/abs/2104.06492}{{\ttfamily 2104.06492}}].

\bibitem{Felder:2017rgg}
G.~Felder and M.~M\"uller-Lennert, \emph{{Analyticity of Nekrasov Partition
  Functions}}, \href{https://doi.org/10.1007/s00220-018-3270-1}{\emph{Commun.
  Math. Phys.} {\bfseries 364} (2018) 683}
  [\href{https://arxiv.org/abs/1709.05232}{{\ttfamily 1709.05232}}].

\bibitem{Kim:2021gyj}
H.-C.~Kim, M.~Kim and S.-S.~Kim, \emph{{5d/6d Wilson loops from blowups}},
  \href{https://doi.org/10.1007/JHEP08(2021)131}{\emph{JHEP} {\bfseries 08}
  (2021) 131} [\href{https://arxiv.org/abs/2106.04731}{{\ttfamily
  2106.04731}}].

\bibitem{Manschot:2019pog}
J.~Manschot, G.W.~Moore and X.~Zhang, \emph{{Effective gravitational couplings
  of four-dimensional $ \mathcal{N} $ = 2 supersymmetric gauge theories}},
  \href{https://doi.org/10.1007/JHEP06(2020)150}{\emph{JHEP} {\bfseries 06}
  (2020) 150} [\href{https://arxiv.org/abs/1912.04091}{{\ttfamily
  1912.04091}}].

\bibitem{Xie:2012hs}
D.~Xie, \emph{{General Argyres-Douglas Theory}},
  \href{https://doi.org/10.1007/JHEP01(2013)100}{\emph{JHEP} {\bfseries 01}
  (2013) 100} [\href{https://arxiv.org/abs/1204.2270}{{\ttfamily 1204.2270}}].

\bibitem{Bonelli:2011aa}
G.~Bonelli, K.~Maruyoshi and A.~Tanzini, \emph{{Wild Quiver Gauge Theories}},
  \href{https://doi.org/10.1007/JHEP02(2012)031}{\emph{JHEP} {\bfseries 02}
  (2012) 031} [\href{https://arxiv.org/abs/1112.1691}{{\ttfamily 1112.1691}}].

\bibitem{Bershtein:2018srt}
M.~Bershtein, P.~Gavrylenko and A.~Marshakov, \emph{{Cluster Toda chains and
  Nekrasov functions}},
  \href{https://doi.org/10.1134/S0040577919020016}{\emph{Theor. Math. Phys.}
  {\bfseries 198} (2019) 157}
  [\href{https://arxiv.org/abs/1804.10145}{{\ttfamily 1804.10145}}].

\bibitem{Nekrasov:2002qd}
N.A.~Nekrasov, \emph{{Seiberg-Witten prepotential from instanton counting}},
  \href{https://doi.org/10.4310/ATMP.2003.v7.n5.a4}{\emph{Adv. Theor. Math.
  Phys.} {\bfseries 7} (2003) 831}
  [\href{https://arxiv.org/abs/hep-th/0206161}{{\ttfamily hep-th/0206161}}].

\bibitem{Nekrasov:1996cz}
N.~Nekrasov, \emph{{Five dimensional gauge theories and relativistic integrable
  systems}}, \href{https://doi.org/10.1016/S0550-3213(98)00436-2}{\emph{Nucl.
  Phys. B} {\bfseries 531} (1998) 323}
  [\href{https://arxiv.org/abs/hep-th/9609219}{{\ttfamily hep-th/9609219}}].

\bibitem{Qiu:2016dyj}
J.~Qiu and M.~Zabzine, \emph{{Review of localization for 5d supersymmetric
  gauge theories}}, \href{https://doi.org/10.1088/1751-8121/aa5ef0}{\emph{J.
  Phys. A} {\bfseries 50} (2017) 443014}
  [\href{https://arxiv.org/abs/1608.02966}{{\ttfamily 1608.02966}}].

\bibitem{Baulieu:1997nj}
L.~Baulieu, A.~Losev and N.~Nekrasov, \emph{{Chern-Simons and twisted
  supersymmetry in various dimensions}},
  \href{https://doi.org/10.1016/S0550-3213(98)00096-0}{\emph{Nucl. Phys. B}
  {\bfseries 522} (1998) 82}
  [\href{https://arxiv.org/abs/hep-th/9707174}{{\ttfamily hep-th/9707174}}].

\bibitem{Bershtein:2015xfa}
M.~Bershtein, G.~Bonelli, M.~Ronzani and A.~Tanzini, \emph{{Exact results for $
  \mathcal{N} $ = 2 supersymmetric gauge theories on compact toric manifolds
  and equivariant Donaldson invariants}},
  \href{https://doi.org/10.1007/JHEP07(2016)023}{\emph{JHEP} {\bfseries 07}
  (2016) 023} [\href{https://arxiv.org/abs/1509.00267}{{\ttfamily
  1509.00267}}].

\bibitem{Nekrasov:2020qcq}
N.~Nekrasov, \emph{{Blowups in BPS/CFT Correspondence, and Painlev\'e VI}},
  \href{https://doi.org/10.1007/s00023-023-01301-5}{\emph{Annales Henri
  Poincare} {\bfseries 25} (2024) 1123}
  [\href{https://arxiv.org/abs/2007.03646}{{\ttfamily 2007.03646}}].

\bibitem{Nakajima:2005fg}
H.~Nakajima and K.~Yoshioka, \emph{{Instanton counting on blowup. II.
  K-theoretic partition function}},
  \href{https://arxiv.org/abs/math/0505553}{{\ttfamily math/0505553}}.

\bibitem{Gottsche:2006bm}
L.~Gottsche, H.~Nakajima and K.~Yoshioka, \emph{{K-theoretic Donaldson
  invariants via instanton counting}},
  \href{https://doi.org/10.4310/PAMQ.2009.v5.n3.a5}{\emph{Pure Appl. Math.
  Quart.} {\bfseries 5} (2009) 1029}
  [\href{https://arxiv.org/abs/math/0611945}{{\ttfamily math/0611945}}].

\bibitem{Shchechkin:2020ryb}
A.~Shchechkin, \emph{{Blowup relations on $\mathbb{C}^2/\mathbb{Z}_2$ from
  Nakajima\textendash{}Yoshioka blowup relations}},
  \href{https://doi.org/10.1134/S0040577921020070}{\emph{Teor. Mat. Fiz.}
  {\bfseries 206} (2021) 225}
  [\href{https://arxiv.org/abs/2006.08582}{{\ttfamily 2006.08582}}].

\bibitem{Kim:2025qaf}
H.-C.~Kim, M.~Kim, S.-S.~Kim, K.~Lee and X.~Wang, \emph{{Probing Quantum Curves
  and Transitions in 5d SQFTs via Defects and Blowup Equations}},
  \href{https://arxiv.org/abs/2503.15591}{{\ttfamily 2503.15591}}.

\bibitem{Huang:2022hdo}
M.-x.~Huang, K.~Lee and X.~Wang, \emph{{Topological strings and Wilson loops}},
  \href{https://doi.org/10.1007/JHEP08(2022)207}{\emph{JHEP} {\bfseries 08}
  (2022) 207} [\href{https://arxiv.org/abs/2205.02366}{{\ttfamily
  2205.02366}}].

\bibitem{fomin2001laurentphenomenon}
S.~Fomin and A.~Zelevinsky, \emph{The laurent phenomenon},
  \href{https://doi.org/https://doi.org/10.1006/aama.2001.0770}{\emph{Advances
  in Applied Mathematics} {\bfseries 28} (2002) 119}
  [\href{https://arxiv.org/abs/math/0104241}{{\ttfamily math/0104241}}].

\bibitem{Fomin:2016caz}
S.~Fomin, L.~Williams and A.~Zelevinsky, \emph{{Introduction to Cluster
  Algebras. Chapters 1-3}},  \href{https://arxiv.org/abs/1608.05735}{{\ttfamily
  1608.05735}}.

\bibitem{Bershtein:2016uov}
M.A.~Bershtein and A.I.~Shchechkin, \emph{{Backlund transformation of Painleve
  III($D_8$) tau function}},
  \href{https://doi.org/10.1088/1751-8121/aa59c9}{\emph{J. Phys. A} {\bfseries
  50} (2017) 115205} [\href{https://arxiv.org/abs/1608.02568}{{\ttfamily
  1608.02568}}].

\end{thebibliography}\endgroup
\end{document}